\documentclass[sn-mathphys-num]{sn-jnl}

\usepackage{graphicx}%
\usepackage{multirow}%
\usepackage{amsmath,amssymb,amsfonts}%
\usepackage{amsthm}%
\usepackage{mathrsfs}%
\usepackage[title]{appendix}%
\usepackage{xcolor}%
\usepackage{textcomp}%
\usepackage{manyfoot}%
\usepackage{booktabs}%
\usepackage{algorithm}%
\usepackage[algo2e]{algorithm2e}%
\usepackage{algorithmicx}%
\usepackage{algpseudocode}%
\usepackage{listings}%
\usepackage{tikz}%
\usepackage{caption}%
\usepackage{subcaption}%
\usepackage{ulem}%

\begin{document}

\title[Article Title]{A computationally efficient procedure for combining ecological datasets by means of sequential consensus inference}

\author*[1]{\fnm{Mario} \sur{Figueira}}\email{Mario.Figueira@uv.es}
\author[1]{\fnm{David} \sur{Conesa}}
\author[1]{\fnm{Antonio} \sur{López-Quílez}}
\author[2]{\fnm{Iosu} \sur{Paradinas}}

\affil*[1]{\orgdiv{Department of Statistics and Operations Research}, \orgname{University of Valencia}, \orgaddress{Carrer del Dr. Moliner, 50, 46100 Burjassot, Valencia, Spain}}

\affil[2]{AZTI, Txatxarramendi Ugartea z/g, 48395 Sukarrieta, Spain}

\abstract{Combining data has become an indispensable tool for managing the current diversity and abundance of data. But, as data complexity and data volume swell, the computational demands of previously proposed models for combining data escalate proportionally, posing a significant challenge to practical implementation. This study presents a sequential consensus Bayesian inference procedure that allows for a flexible definition of models, aiming to emulate the versatility of integrated models while significantly reducing their computational cost. 
The method is based on updating the distribution of the fixed effects and hyperparameters from their marginal posterior distribution throughout a sequential inference procedure, and performing a consensus on the random effects after the sequential inference is completed. The applicability, together with its strengths and limitations, is outlined in the methodological description of the procedure.
The sequential consensus method is presented in two distinct algorithms. The first algorithm performs a sequential updating and consensus from the stored values of the marginal or joint posterior distribution of the random effects. The second algorithm performs an extra step, addressing the deficiencies that may arise when the model partition does not share the whole latent field.
The performance of the procedure is shown by three different examples -one simulated and two with real data- intending to expose its strengths and limitations.
}


\maketitle

\section{Introduction}\label{sec:Introduction}

The field of ecology is undergoing a transformation fuelled by the availability of diverse and abundant datasets. Historically, ecological research has relied on limited data streams, often constrained by logistical challenges and disciplinary boundaries. However, recent advancements in technology and the proliferation of interdisciplinary collaborations have ushered in an era of data-driven ecology. In isolation, each dataset offers estimations of the ecological process under investigation; however, their integration may yield more refined estimations resulting in reduced uncertainty estimates \citep{Fletcher_CombiningData_2019, Rufener_Bridging_2021, Alglave_CombiningIndPref_2022, Paradinas_ISDM_2023}. Different methods for combining data vary in their capacity to address sampling biases, establish linkages between divergent response variables across datasets, and effectively manage inherent uncertainties within the data \citep{Fletcher_CombiningData_2019}.

The most straightforward method is {\sl data pooling}, which involves aggregating data without explicitly accounting for their diverse sources and associated sampling biases. This approach assumes uniformity in the nature of the response variable across all datasets. In cases where the data sources differ in type, transformation of one dataset is necessary, potentially leading to information loss \cite{Isaac_DataIntegration_2020, Fithian_Bias_2015}. For example, this transformation could involve converting abundance data into presence-absence or presence-only data, thereby degrading the quality of the information in the data.

Another method is {\sl ensemble modelling}, where multiple diverse models are created to predict a unique outcome, either by using many different individual models to the same dataset or by using one modelling set up to different datasets \citep{Araujo_Ensemble_2007, Nisbet_HandbookStatisticalAnalysis_2018}. However, formal integration of parameter estimates may pose challenges unless both datasets exhibit similar resolutions \citep{Fletcher_CombiningData_2019} and adhere to consistent response variable types.

A more formal approach for combining data involves modelling various datasets simultaneously. This approach is known as {\sl integrated modelling}, where the model explicitly addresses the differences in their sampling processes. The strength of integrated models stem in their ability to combine information from diverse datasets, enabling the estimation of shared parameters across models through joint-likelihood procedures \citep{Fletcher_SpatialEcology_2019, Alglave_CombiningIndPref_2022, Rufener_Bridging_2021, Paradinas_ISDM_2023}. Unlike data pooling and ensemble modelling techniques, integrated models offer a formal framework for combining different types of data and sampling procedures. As the complexity and scale of data increase, the computational demands of integrated models escalate proportionally, posing significant challenges in practical implementation.

An alternative and faster approach is to combine information using {\sl Sequential inference}, which is based on recursive Bayesian inference \citep{Hooten_RecursiveBayesianInference_2021} and sequential approaches \citep{Doucet_SequentialMonteCarlo_2001, Scott_ConsensusMonteCarlo_2016, Scott_ConsensusDistributedComputation_2017}, as implemented in other scientific fields such as neurology \citep{Kording_BayesianIntegration_2004}, biometry \citep{Zigler_ModelFeedbackBiometric_2013}, machine learning \citep{Ngunyen_BayesianAbstentionFeedbacks_2022} or quantum physics \citep{Brakhane_BayesianFeedbackQuantumPhysics_2012}. In this case, information is sequentially incorporated by means of the posterior distributions of the parameters and hyperparameters of a model as the new prior distributions of the next model. This procedure is repeated until all datasets have been fitted in a sequence.

Nevertheless, although theoretically performing sequential inference produces the exact same results than those from a complete and simultaneous inference approach, implementing this procedure in models with complex latent structures (such as spatio-temporal models) is not straightforward \citep{Figueira_BayesianFeedback_2023}. This limitation could be overcame by an approximate sequential inference procedure implemented within the Integrated Nested Laplace Approximation (INLA) \citep{Rue_INLA_2009} framework, by exploiting the methodological underpinnings of the Laplace approximation and the Latent Gaussian Model (LGM) structure. As shown in this paper, this method can substantially reduce the computational burden of integrated models while maintaining a high degree of fidelity to the underlying data dynamics.

In particular, the objective of this study is to propose a framework for combining multiple sources of information, whether derived from different types of data or different sampling structures. The proposed method, termed sequential consensus by its similarity to other sequential approaches \citep{Doucet_SequentialMonteCarlo_2001, Scott_ConsensusDistributedComputation_2017}, improve upon this by introducing a new layer to overcome the limitation of not sharing information about the latent field random effects and aims to offer the same versatility as previous models while reducing computational costs by analysing the various sources of information separately. This is achieved by integrating part of the model information sequentially, specifically pertaining to fixed parameters and hyperparameters, and subsequently aggregating the remaining information related to the random effects of the latent field by consensus. We compare our method across simulated and real scenarios with the results obtained from a complete and simultaneous modelling, which serve as the gold standard for evaluating the performance of our proposed algorithm. Results show that both methodologies produce very similar or indistinguishable results, allowing us to proceed with the analysis in parts to considerably reduce the computational cost.

After this Introduction, in Section 2 we briefly review spatio-temporal models, our selection among the long list of models with complex structures where incorporating information can result in large computational burden. The section also includes a brief review of the INLA approach. Section 3 describes in detail our proposal to perform a consensus sequential approach, while in Section 4 we present the results of applying our approach in two real examples along with a simulated example. We conclude in Section 5.

\section{Inference and prediction in spatio-temporal modelling}

Complex spatio-temporal models could be the quintessential example of computational burden \citep{Banerjee_HierarchicalSpatialData_2015, Paradinas_SpatioTemporal_2017}. In such cases, sequential inference can be particularly advantageous, potentially reducing computation time and making the analysis feasible. The computational challenge of spatio-temporal models is amplified as multiple datasets are considered, each contributing additional layers of information and nuance. Therefore, the optimisation of computational efficiency in spatio-temporal modelling is a great example for fully tapping the potential of sequential modelling in addressing complex real-world phenomena spanning domains such as species distribution models, climate science, public health, and a large etcetera.

At the forefront of this field lie geostatistics. Grounded in the principles of spatial dependence and variability, geostatistics provides a robust framework for characterising and predicting spatial processes through statistical inference. Geostatistics yields reliable estimates when applied across a randomly selected samples. However, when samples are preferentially gathered (such as in citizen science data), it becomes crucial to address this inherent dependence in the analysis. Moreover, when temporal dependence appears in the data, the spatial domain shall be expanded to the spatio-temporal one, enabling to evaluate these effects jointly. Further, incorporating various sources of information that account for the peculiarities of each of them (e.g. sampling designs) can be achieved through integrated modelling. All these models can be implemented in the well-known and extensively used \texttt{R-INLA} software.

\subsection{Modelling geostatistical data}

Geostatistical models assume that data are generated from a continuous spatial process $\{y(\mathbf{s}), \mathbf{s}\in\mathcal{D}\}$, where $y(\mathbf{s})$ are the geostatistical or point-referenced data realisations and $\mathbf{s}$ is a spatial index varying continuously in the spatial domain $\mathcal{D}$ \citep{Diggle_Geostatistics_1998, Diggle_ModelBasedGeostatistics_2007, Banerjee_HierarchicalSpatialData_2015}. 

In general, a geostatistical model can be formed by an intercept $\beta_0$, a set of linear effects $\boldsymbol\beta$ for a matrix covariates $\mathbf{A}$, non linear random effects $f_k(z_i)$ and some spatial $u_s$ random effects:
\begin{equation}
\begin{array}{c}
     y_i \mid \eta_i, \boldsymbol\theta \sim \ell(y_i \mid \mu_i, \boldsymbol\theta), \\
     g(\mu_i) = \beta_0 + \boldsymbol\beta \mathbf{A}_{i} + \sum_{j=1}^J f_j(z_{ij}) + u_i,
\end{array}
\label{eq:geostatistical_model}
\end{equation}
where $\mu_i$ is the mean of the likelihood of the data ($\ell$), $\eta_i$ is the linear predictor $g(\mu_i) = \eta_i$, $\boldsymbol\theta$ represents the hyperparameters and $u_i$ is a spatial effect. 

This structure is a good approximation to analyse data with a smooth spatial \citep{Nychka_MultiresolutionGaussianProcess_2015} or space-time dependence \citep{Blangiardo_SpatioTemporalINLA_2013} that is not explained by explanatory variables, and it is extensively used in species distribution models \citep{Cosandey_SpatioTemporalArtic_2015, Paradinas_SpatioTemporal_2015} as in many other fields \citep{Blangiardo_SpatioTemporalINLA_2013, Jingyi_EnvironmentalINLA_2017}.

\subsection{Preferential model}

The underlying assumption in a geostatistical model is that locations where data are observed $\mathbf{s}=\{s_1,..,s_n\}$ are independent from the marks (values observed) at those sampling locations. Indeed, this is a very restrictive assumption that sometimes (due to time and financial constraints) does not hold \citep{Diggle_Preferential_2010}: air-quality monitoring sites are located in those places where one can think that the air quality is lower; observers tend to search in areas where they expect (there is higher probability) to find a specific species, etc. 

In a preferential model, the locations and their corresponding marks are generated by two different but related processes: a spatial point process that generates the observed spatial locations, and a model for the quantities observed at each location.

The observed locations $\mathbf{s}=\{s_1,...,s_n\}$ are a realisation of a non-homogeneous Poisson process, namely a log-Gaussian Cox process (LGCP), where the intensity function $\lambda(s)$ is modelled as a random field that follows a log-Gaussian distribution. More precisely, a LGCP is defined in a bounded region $\Omega \subset \mathbb{R}^2$, where the number of points within a subregion $D \subset \Omega$ is Poisson-distributed with a expected value $\Lambda(D)=\int_{D}\lambda(s)ds$, where $\lambda(s)$ is the intensity function of the point process with a spatial structure. 

The likelihood of an LGCP, given the intensity function and the {\sl marked} point pattern (the set of samples) $\mathbf{s}=\{s_1,...,s_n\}$ is
\begin{equation}
    \pi(y(\mathbf{s})\mid \lambda) = \exp\left[ |\Omega|- \int_{\Omega}\lambda(s)ds \right]\prod_{s_i\in \mathbf{s}}\lambda(s_i),
\end{equation}
where the form of the likelihood can be particularly difficult to deal with analytically, but can be solved by computational numerical methods \citep{Illian_ToolboxLGCP_2012, Diggle_LogGaussianCoxProcesses_2013, Simpson_GoingOffGrid_2016, Sorbye_priorLGCP_2018}. Thus, by modelling the log-intensity, we can define a versatile class of point processes. 

As above mentioned, a preferential model is completed by defining the likelihood for the model of the marks. Sharing information between the linear predictor for the log-intensity and the linear predictor for the marks enables us to establish a preferential model. This joint model can effectively incorporate the sampling process while capturing relationships between the sampling and the underlying process for the marks, usually the spatial or spatio-temporal structure:
\begin{equation}
\begin{array}{c}
     \displaystyle g(\mu_i) = \beta_0 + \boldsymbol\beta\mathbf{A}_i + \sum_{j=1}^Jf_j(z_{ij}) + u_i, \\
     \displaystyle \log(\lambda(s_i)) = \beta^*_0 + \boldsymbol\beta^*\mathbf{A}^*_i + \sum_{j=1}^Jf^*_j(z^*_{ij}) + \alpha \cdot  u_i + u^*_i,
\end{array}
\end{equation}
where $\mu_i$ is the mean of the distribution of the marks and the elements in the linear predictor for $g(\mu_i)$ are similar to those of the geostatistical model in Equation (\ref{eq:geostatistical_model}). The components for the log-intensity can mirror those of the linear predictor for $g(\mu_i)$, encompassing identical fixed effects associated with the same explanatory variables and the same random effect structures $\{\beta_0^*,\boldsymbol\beta^*, \mathbf{A}^*, f_k^*(z^*_{ik})\}$. The preferential element is the coefficient $\alpha$, as it allows to share the spatial or the spatio-temporal effect between the log-intensity and the linear predictor of the marks. Finally, is is also possible to introduce a spatial or spatio-temporal term defined specifically to the point process, $u^*_i$.

It is worth noting that in addition to the shared components mentioned above, any other elements of the latent field can be shared between different linear predictors \citep{Paradinas_SpatioTemporal_2017, Krainski_AdvancedSPDE_2018}. For instance, one might include a random effect like first-order or second-order random walk associated to some covariate, a purely temporal trend or any other relevant process. Each of these shared components can be assigned different scaling parameters, or they can be uniform across several predictors, or even set equal to one. This flexibility allows for the joint modelling of effects using various types of data.

In short, the preferential model is a joint model where specific components are shared between the linear predictors of the point process and the geostatistical process. These shared terms, when scaled by a parameter $\alpha_k$, determine whether preferentiality is positive if $\alpha_k$ is positive, or negative if the parameter is negative \citep{Diggle_Preferential_2010}, with regard to the $k$-th shared component.

\subsection{Spatio-temporal model}

In the previous models we have focused on the spatial component and the association between the point pattern of sampling and the values of the marks. However, these models can be extended by incorporating temporal terms in conjunction with spatial terms, as hinted at earlier. In this subsection, we will provide a brief overview of the different structures that can be encountered in spatio-temporal modelling.

Spatio-temporal models allow the simultaneous assessment of spatial and temporal dynamics \citep{Krainski_AdvancedSPDE_2018, Wikle_SpatioTemporalStatistics_2019}, and have been widely used in various fields: in epidemiology \citep{Moraga_MalariaINLA_2021}; in species distribution models \citep{Paradinas_SpatioTemporal_2015, Yuan_PointProcessDistanceSampling_2017}; in models to assess animal movement and habitat preference \citep{Arce_SpatialVariationStepSelection_2023, Michelot_StepSelectionAnalysis_2024}; in econometric models \citep{Virgilio_SpatialEconometricsINLA_2021}; and in many others. 

A comprehensive overview of various spatio-temporal structures is provided in \citep{Blangiardo_SpatioTemporalINLA_2013, Paradinas_SpatioTemporal_2017} by decomposing the spatio-temporal term into two distinct components: a pure temporal trend and a pure spatial effect. Depending on their structural characteristics, the following types of structures can be identified: (a) opportunistic spatial distribution, where $u_{st} = w_{st}$, with $w_t \sim \mathcal{N}(\mathbf{0}, \mathbf{Q})$ representing different spatial realisations with consistent hyperparameters across time (essentially replicas of the spatial effect over time); (b) Persistent spatial distribution with random intensity changes over time, expressed as $u_{st} = w_{st} + v_t$, where the spatio-temporal effect is segregated into a spatial trend replicated across time and an independent and identically distributed (i.i.d.) temporal effect $v_t \sim \mathcal{N}(0, \tau_v)$; (c) persistent spatial distribution with temporal intensity trend, where $u_{st}= w_s + v_{t}$, implying the same spatial effect over time and a temporal trend defined by a structured random effect (\textit{rw1}, \textit{rw2}, \textit{ar1}, ...); and finally a (d) progressive spatio-temporal distribution, where the spatio-temporal effect is decomposed as $u_{st} = w_{st} + v_{t}$, with the spatial effect is replicated across time, alongside a structured temporal trend.

In addition to being able to evaluate spatio-temporal models by decomposing them into two terms, one purely spatial and one purely temporal, it is possible to make a similar synthesis but considering different structures for spatial and temporal interaction components. \cite{KnorrHeld_SpaceTime_2000} proposed four general structures to define spatio-temporal interaction models, in which the precision matrix of the spatio-temporal effect is constructed by means of the Kronecker product of the precision matrix of the spatial component by the precision matrix of the temporal component $\mathbf{Q}_{st}=\mathbf{Q}_s\otimes\mathbf{Q}_t$ \citep{Clayton_GLMM_1996}. According to this approach, four types of interaction emerge naturally by crossing unstructured and structured precision matrices for space and time. 

Finally, it is also possible to construct non-separable spatio-temporal interaction models \citep{Lindgren_diffusionbased_2024}, i.e. those whose precision matrix can not be decomposed through the Kronecker product of a precision matrix associated to the spatial structure and another one associated to the temporal structure, that is, $\mathbf{Q}_{st} \neq \mathbf{Q}_s \otimes \mathbf{Q}_t$.

\subsection{Integrated models}

Integrated models allow different sources of information to be combined by sharing components in the same way as described above for preferential models. These models can combine information from samples with different structures \citep{Alglave_CombiningIndPref_2022}, such as completely random samples, stratified random samples or preferential samples. But they can also be used to combine information from different types of data on variables that share components in the latent structure \citep{Koshkina_ISDMimperfectdetection_2017, Fletcher_CombiningData_2019, Jung_IntegratedSpeciesDistributionModels_2023, Paradinas_ISDM_2023}.

The following two situations in ecological and environmental contexts respectively could be handled with integrated models: combining the number of catches of a species with other information on the abundance of the same species; and combining information on the presence/absence of a toxin along a river with other measures of its concentration at different locations. In both cases, by combining the two sources of information in a joint model, it becomes possible to analyse both variables simultaneously and to use common elements of the latent field for a more accurate and robust estimation.

In both examples, we have two different likelihoods whose linear predictors can be connected by either scaling a shared component with a parameter $\alpha$ or by sharing the effect without scaling it:
\begin{equation}
\begin{array}{c}
     y_{1i} \mid \eta_{1i}, \boldsymbol\theta_1 \sim \ell_1(y_{1i}\mid \eta_{1i}, \boldsymbol\theta_1), \\
     y_{2j} \mid \eta_{2j}, \boldsymbol\theta_2 \sim \ell_2(y_{2j}\mid \eta_{2j}, \boldsymbol\theta_2), \\
     g_1(\mu_{1i}) = \eta_{1i} = \beta_{10} + \mathbf{A}_{1i}\boldsymbol\beta_1 + u(s_{i}), \\
     g_2(\mu_{2j}) = \eta_{2j} = \beta_{20} + \mathbf{A}_{2j}\boldsymbol\beta_2 + \alpha\cdot u(s_{j}), \\
\end{array}
\end{equation}
where $\ell_1$ and $\ell_2$ are different likelihood functions for $\mathbf{y}_{1}$ and $\mathbf{y}_{2}$, and $\boldsymbol{\theta}_1$ and $\boldsymbol{\theta}_2$ are the set of hyperparameters associated with each likelihood. Additionally, there are two different intercepts $\{\beta_{10}, \beta_{20}\}$ and $\{\mathbf{A}_1, \mathbf{A}_2\}$ denoting sets of explanatory variables with their linear coefficients $\{\boldsymbol\beta_1, \boldsymbol\beta_2\}$. Both linear predictors shared the spatial component $\mathbf{u}(s)$, scaled by a $\alpha$ parameter in the second linear predictor. This illustrates the flexibility of the integrated modelling approach to analyse multiple sources of information together.

\subsection{INLA}

The methodology proposed in this paper for combining information from different sources has been implemented within the framework of the Integrated Nested Laplace Approximation \citep{Rue_INLA_2009} in the \texttt{R-INLA} software \citep{Martins_BayesianComputingINLA_2013}, although it could easily be adapted to other approaches such as Monte Carlo or Markov Chain Monte Carlo methods.

INLA is a deterministic approximation approach deeply rooted in Gaussian Markov Random Field (GMRF) theory \citep{Rue_GMRF_2005} for Bayesian inference \citep{Rue_INLA_2009, VanNiekert_NewINLA_2023}. Its essence lies in the efficient computation of marginal posterior distributions within a wide class of models known as latent Gaussian models. INLA achieves this by exploiting the conditional independence properties inherent in GMRFs, allowing complex data structures to be modelled and analysed in a computationally efficient manner. Through a series of nested approximations, INLA estimates the marginal posterior distributions of the model parameters, allowing the calculation of marginal likelihoods and standard goodness-of-fit metrics such as Deviance Information Criterion (DIC) \citep{Spiegelhalter_DIC_2002}, Watanabe-Akaike Information Criterion (WAIC) \citep{Watanabe_WAIC_2013} and Conditional Predictive Ordinates (CPO) \citep{Pettit_CPO_1990}, all of which are essential for model evaluation and comparison. 

In addition, the incorporation of the SPDE-FEM approach in \texttt{R-INLA} allows spatial modelling, providing a Matérn covariance function and allowing direct calculation of its precision matrix. This allows more complex spatial and spatio-temporal structures to be defined as non-stationary models or non separable space-time models \citep{Lindgren_ExplicitLinkSPDE_2011, Krainski_AdvancedSPDE_2018, Bakka_SpatialBarriersINLA_2019, Siden_3DSpatialMatern_2021, Lindgren_diffusionbased_2024}.

It is worth noting that INLA is well integrated across different scientific disciplines, including ecology \citep{Pennino_AccountingPreferential_2019, Nerea_SpinetailSPDE_2020, Paradinas_ISDM_2023}, econometrics \citep{Bivand_SpatialEconometricsINLA_2014, Virgilio_SpatialEconometricsINLA_2021}, epidemiology \citep{Moraga_MalariaINLA_2021}, environmental science \citep{Jingyi_EnvironmentalINLA_2017}, and others. This clearly indicates that the implementation of the proposed methodology has a dual significance: to exploit the fundamental principles of the INLA approach and to benefit from its widespread use in different scientific fields.

\section{Methodology}

This section presents a framework for combining the information obtained from different datasets or set of datasets. This approach reduces the computational burden by sequentially updating the information of those datasets or set of datasets. The methodology focuses on obtaining the posterior distributions of the latent field and hyperparameters taking advantage of working with Gaussian latent fields within the INLA approach.

\subsection{Sequential inference}

The underlying idea of a sequential analysis of a dataset, or a set of datasets, is to divide it into $n$ subsets $\mathbf{y}=\{\mathbf{y}_1, ..., \mathbf{y}_n\}$ and fit the model for each subset by sequentially updating the joint prior distribution of the latent field and hyperparameters of each model with the posterior distributions of the previous model. Sequential inference relies in the use of the conditional independence property $\pi(y_i \mid y_{-i}, \eta_i, \boldsymbol\theta) = \pi(y_i \mid \eta_i, \boldsymbol\theta)$, given the linear predictor $\eta_i$ and the hyperparameters $\boldsymbol\theta$. We can decompose the joint posterior distribution of the latent field $\mathbf{x}$ and the hyperparameters $\boldsymbol\theta$ as:
\begin{equation}
\begin{array}{rcl}
\pi(\mathbf{x},\boldsymbol\theta \mid \mathbf{y}) & \propto & \pi(\mathbf{y} \mid \mathbf{x},\boldsymbol\theta)\; \pi(\mathbf{x},\boldsymbol\theta), \\
 & = & \prod_{i=1}^n\pi(\mathbf{y}_i \mid \mathbf{x},\boldsymbol\theta)\;  \pi(\mathbf{x},\boldsymbol\theta),
\end{array}
\end{equation}
where each subset $\mathbf{y}_i$ is related to one model and $\pi(\mathbf{x},\boldsymbol\theta \mid \mathbf{y}_1,...,\mathbf{y}_i)$ stands for the joint posterior distribution of the latent field and hyperparameters until the $i$-th subset.

INLA focuses on the marginals of the latent field and the hyperparameters, thus it is not possible to obtain the joint distribution of the latent field and hyperparameters $\pi(\mathbf{x},\boldsymbol\theta \mid \mathbf{y})$. Still, in order to perform a sequential inference, instead of sharing the joint posterior distribution, it is possible to share information only between the different common fixed parameters and hyperparameters between the models \citep{Figueira_BayesianFeedback_2023}.

In particular, using the reasoning for the joint posterior distribution, and in line with \citep{Figueira_BayesianFeedback_2023}, we propose the following approximation to calculate the marginal posteriors of the fixed effects $\boldsymbol\beta = \{\beta_1, ..., \beta_K\}$, given a partition of the dataset $\mathbf{y}=\{\mathbf{y}_1,..., \mathbf{y}_n\}$:
\begin{equation}
\begin{array}{rcl}
    \displaystyle \pi(\beta_k \mid \mathbf{y}) & \propto & \pi(\mathbf{y} \mid \beta_k) \; \pi(\beta_k), \\ 
    & = & \displaystyle \prod_{i=1}^n \pi(\mathbf{y}_i \mid \beta_k)\; \pi(\beta_k), \\
    & = & \displaystyle \prod_{i=2}^n \pi(\mathbf{y}_i \mid \beta_k)\; \pi(\mathbf{y}_1 \mid \beta_k)\; \pi(\beta_k), \\
    & \propto & \displaystyle \prod_{i=2}^n \pi(\mathbf{y}_i \mid \beta_k)\; \pi(\beta_k \mid \mathbf{y}_1)
\end{array}
\end{equation}
where the posterior of the step $i-1$, $\pi(\beta_k \mid \cup_{j=1}^{i-1}\mathbf{y}_j)$, is used as the prior for the following step $i$ for the fixed effect $\beta_k$.

With respect to the marginal posterior of the hyperparameters, $\boldsymbol\theta = (\theta_1, \ldots, \theta_K)$, we propose a similar approximation:
\begin{equation}
\begin{array}{rcl}
    \displaystyle \pi(\theta_k \mid \mathbf{y}) & \propto & \pi(\mathbf{y} \mid \theta_k)\; \pi(\theta_k) \\
    & \propto & \displaystyle \prod_{i=2}^n \pi(\mathbf{y}_i \mid \theta_k)\; \pi(\theta_k \mid \mathbf{y}_1), \\
\end{array}
\end{equation}
where again the marginal posterior of $\theta_k$ is used as the prior for the following step.

It is worth noting that this sequential inference procedure does not provide an update of the random effects of the latent field. To overcome this deficiency, in what follows we present two different consensus procedures for updating the random effects by combining the information of the latent field random effects between the different modelling outputs along the data subsets. The first one is based on marginal weighted averages, while the second one focuses on the distribution of each random effect.

\subsection{Marginal weighted averages}

Our first proposal for combining the information of the latent field random effects is based on averaging their marginal distributions. Given that each node in a latent Gaussian field is a random variable following a normal distribution $x_i \sim \mathcal{N}(\mu_i, \tau_i)$, with mean $\mu_i$ and precision $\tau_i$, it is possible to combine information from a random effect structure $\mathcal{X}=\{x_1,..., x_k\}$ that is used in several models into which the dataset has been divided. 

In particular, our proposal is to approximate the posterior of the marginals for each node $x_i \in \mathcal{X}$ by a weighted averaging along the different $n$ models in which $\mathcal{X}$ appears \citep{Huang_SamplingFB_2005, Scott_ConsensusMonteCarlo_2016}:
\begin{equation}
    x_i \approx \sum_{j=1}^n w_{ij} x_{ij},
\label{eq:weighted_averages}
\end{equation}
where $x_{ij} \sim \mathcal{N}(\mu_{ij}, \tau_{ij})$ represents the marginal random variable for $i$-th node from the $j$-th model with mean $\mu_{ij}$ and precision $\tau_{ij}$; and $w_{ij}=\tau_{ij}/\sum_{l=1}^n\tau_{il}$ are the optimal weights for Gaussian random variables \citep{Huang_SamplingFB_2005, Scott_ConsensusMonteCarlo_2016}, such that $\sum_{j=1}^n w_{ij}=1$. As a result, each node $x_i$ is Gaussian distributed with mean and precision:
\begin{equation}
\begin{array}{c}
    \mu_i = \displaystyle \sum_{j=1}^n w_{ij}\mu_{ij} \,, \\
    \tau_i = \displaystyle \left(\sum_{j=1}^n w^2_{ij}/\tau_{ij}\right)^{-1} = \sum_{l=1}^n\tau_{il} \,.
\end{array}
\end{equation}

Performing this approximation along the whole set of nodes $\{x_1,...,x_k\}$ related to $\mathcal{X}$, we can combine the information for the latent field random effects. Furthermore, the weighted averaging approach for the marginal distributions of the random effects also allows the use of other weights, $\mathbf{w}_e$ such as $\sum_{i=1}^n w_{ie}=1$, that can be proposed by experts. Note that this expert elicitation of weights mimics the weighted likelihood approach \citep{Fletcher_CombiningData_2019}, allowing in both cases to fit weighted joint models by incorporating several data sources of different quality. The expert weights can be directly used in the Equation (\ref{eq:weighted_averages}), or combined with the optimal weights proposed $w_i=\tau_i/\sum_{j=1}^n\tau_j$. For example, we can redefine the weights as:
\begin{equation}
    w^*_i = \frac{w_{ie}\cdot w_i}{\sum_{j=1}^n w_{ie} \cdot w_i}\,,
\end{equation}
where we blend expert suggested weights with the optimal weights for Gaussian random variables. However, with the introduction of the new weights, the precision for the averaged variable would not be determined by the simple sum $\tau_i = \sum_{j=1}^n\tau_{ij}$, but rather by the more general expression $\tau_i= \left(\sum_{j=1}^n {w^*}^2_j/\tau_{ij}\right)^{-1}$.

It is finally worth noting that using the optimal weights, the distribution of $x_i$ is equivalent to that obtained by calculating the product of the univariate Gaussian densities $\pi(x_i)=\prod_{j=1}^n\pi(x_{ij})$. The following approach generalises this to the multivariate distribution of random effects.

\subsection{Product of multivariate Gaussian densities}

In this second approach, our emphasis lies in integrating comprehensive information regarding the structure of each random effect $\mathcal{X}$, rather than solely focusing on the marginal distribution of each node. This can be done using the properties of multivariate Gaussian distributions that allow us to combine the latent field information obtained in the fitting of each subset of the full dataset \citep{Huang_SamplingFB_2005}. In particular, we can approximate the density of the multivariate posterior distribution $\pi(\mathbf{x}\mid \mathbf{y})$ for a specific random effect as:
\begin{equation}
    \pi(\mathbf{x} \mid \mathbf{y}) \approx \prod_{i=1}^n\pi(\mathbf{x} \mid \mathbf{y}_i) \,, 
\end{equation}
where $\pi(\mathbf{x} \mid \mathbf{y}_i)$ represents the multivariate Gaussian density of the latent field $\mathbf{x}$ with mean $\boldsymbol\mu_i$ and precision matrix $\mathbf{Q}_i$. Consequently, the product $\pi(\mathbf{x} \mid \mathbf{y})$ is another multivariate Gaussian density, denoted as $\mathbf{x} \sim \mathcal{N}(\boldsymbol\mu, \mathbf{Q})$. The corresponding precision matrix and mean of $\pi(\mathbf{x}\mid \mathbf{y})$ can be easily calculated leveraging Gaussian properties:
\begin{equation}
\begin{array}{c}
    \mathbf{Q} = \displaystyle \sum_{i=1}^n \mathbf{Q}_i\,, \\
    \boldsymbol\mu = \displaystyle \mathbf{Q}^{-1}\sum_{i=1}^n\mathbf{Q}_i\boldsymbol\mu_i\, . \\
\end{array}
\end{equation}

Although this approximated multivariate posterior distribution contains all the required information of each latent field random effect, the fact that we can have direct access to its marginals allows us to better describe it. Indeed, we can compute the marginal distribution for each node $x_i \sim \mathcal{N}(\mu_i,\tau_i)$ of the multivariate distribution $\mathbf{x}\sim \mathcal{N}(\boldsymbol\mu, \mathbf{Q})$ related to the random effect $\mathcal{X}$ as:
\begin{equation}
\begin{array}{c}
    \mu_i = \boldsymbol\mu_i\,, \\
    \tau_i = \mathbf{Q}_{ii}\,.
\end{array}
\end{equation}

With this new proposal, we are able to reconstruct the multivariate posterior distributions and the marginal distributions of different random effects (such as spatial effects $\mathcal{X}_s$, temporal effects $\mathcal{X}_t$, or other nonlinear random effects $\mathcal{X}_f$) by combining the information from different models.

The complete consensus sequential framework for the partition $\mathbf{y}=({\mathbf{y}_1,\ldots,\mathbf{y}_n})$, where the posterior marginal distributions of the fixed effects and hyperparameters are used as priors for modelling the next element $\mathbf{y}_i$ of the partition, is summarised in Figure \ref{fig:Scheme_sequentialconsensus}. The figure also represents that once the sequential procedure is completed, the information related to the posterior distributions of the random effects is integrated, either through marginal weighted averages or through the product of multivariate Gaussian densities. Finally, it also shows how when these steps have been carried out, the results are the final posterior distributions given by the sequential consensus Bayesian inference procedure. 

{
\linespread{1.}
\begin{figure}
    \centering
    \begin{tikzpicture}
    \node at  (0,0) [rectangle,draw] (1) {$\everymath={\displaystyle}
    \begin{array}{c}
    \text{\underline{Modelling}} \; \mathbf{y}_1 \\[0.2cm]
    \pi(\boldsymbol\beta\mid\mathbf{y}_1)\\[0.2cm]
    \pi(\mathbf{x}_{-\boldsymbol\beta}\mid\mathbf{y}_1)\\[0.2cm]
    \pi(\boldsymbol\theta\mid\mathbf{y}_1)\\
    \end{array}$};

    \node at  (2.25,0) [rectangle] (prior_to_2) {$\everymath={\displaystyle}
    \begin{array}{c}
    \pi(\boldsymbol\beta\mid\mathbf{y}_1)\\[0.2cm]
    \pi(\boldsymbol\theta\mid\mathbf{y}_1)\\
    \end{array}$};
    
    \node at  (4.5,0) [rectangle,draw] (2) {$\everymath={\displaystyle}
    \begin{array}{c}
    \text{\underline{Modelling}} \; \mathbf{y}_2 \\[0.2cm]
    \pi(\boldsymbol\beta\mid\cup_{i=1}^2\mathbf{y}_i)\\[0.2cm]
    \pi(\mathbf{x}_{-\boldsymbol\beta}\mid\mathbf{y}_2)\\[0.2cm]
    \pi(\boldsymbol\theta\mid\cup_{i=1}^2\mathbf{y}_i)\\
    \end{array}$};

    \node at  (7.1,0) [rectangle] (prior_to_n) {$\everymath={\displaystyle}
    \begin{array}{c}
    \ldots\pi(\boldsymbol\beta\mid\cup_{i=1}^{n-1}\mathbf{y}_i)\\[0.2cm]
    \ldots\pi(\boldsymbol\theta\mid\cup_{i=1}^{n-1}\mathbf{y}_i)\\
    \end{array}$};

    \node at  (9.8,0) [rectangle,draw] (n) {$\everymath={\displaystyle}
    \begin{array}{c}
    \text{\underline{Modelling}} \; \mathbf{y}_n \\[0.2cm]
    \pi(\boldsymbol\beta\mid\cup_{i=1}^n\mathbf{y}_i)\\[0.2cm]
    \pi(\mathbf{x}_{-\boldsymbol\beta}\mid\mathbf{y}_n)\\[0.2cm]
    \pi(\boldsymbol\theta\mid\cup_{i=1}^n\mathbf{y}_i)\\
    \end{array}$};

    \node at  (3,-4.5) [rectangle,draw] (x) {$\everymath={\displaystyle}
    \begin{array}{c}
    \text{\underline{Combining random effect posteriors}} \\[0.2cm]
    1.\;\text{Marginals}\quad x_i \approx \sum_{j=1}^n w_{ij}x_{ij}  \\[0.2cm]
    2.\;\text{Multivariate}\quad \pi(\mathbf{x}_i\mid \mathbf{y})\approx \prod_{j=1}^n \pi(\mathbf{x}_i\mid \mathbf{y}_j)
    \end{array}$};

    \node at  (10.5,-2.25) [rectangle] (posterior_n) {$\everymath={\displaystyle}
    \begin{array}{c}
    \pi(\boldsymbol\beta\mid\mathbf{y})\\[0.2cm]
    \pi(\boldsymbol\theta\mid\mathbf{y})\\
    \end{array}$};    

    \node at  (9.8,-4.5) [rectangle,draw] (result) {$\everymath={\displaystyle}
    \begin{array}{c}
    \text{\underline{Posterior distributions}} \\[0.2cm]
    \pi(\boldsymbol\beta\mid\mathbf{y})\\[0.2cm]
    \pi(\mathbf{x}_{-\boldsymbol\beta}\mid\mathbf{y})\\[0.2cm]
    \pi(\boldsymbol\theta\mid\mathbf{y})\\
    \end{array}$};

    \draw[-to] (1) to [out=0,in=180] (2);
    \draw[-to] (2) to [out=0,in=180] (n);

    \node at  (4.5,2.3) [rectangle] (alg2) {$\everymath={\displaystyle}
    \begin{array}{c}
    \pi(\boldsymbol\beta\mid\mathbf{y}), \; \pi(\boldsymbol\theta\mid\mathbf{y})
    \end{array}$};

    \draw[dotted,-to] (n) to [out=90,in=90,looseness=0.25] (1);
    \draw[dotted,-to] (n) to [out=90,in=90,looseness=0.25] (2);

    \draw[double,-to] (1) to [out=-90,in=90,looseness=0.95] (x);
    \draw[double,-to] (2) to [out=-90,in=90,looseness=1.25] (x);
    \draw[double,-to] (n) to [out=-90,in=90,looseness=0.75] (x);

    \draw[-to] (n) to [out=-90,in=90,looseness=1.5] (result);
    \draw[double,-to] (x) to [out=0,in=180,looseness=0.5] (result);
    \end{tikzpicture}
    \vspace{5mm}
    \caption{Scheme of the sequential consensus framework for the data partition $\mathbf{y}=\{\mathbf{y}_1, \ldots, \mathbf{y}_n\}$: updating in sequence the fixed effects and hyperparameters and performing a consensus for the random effects after the sequential updating.}
    \label{fig:Scheme_sequentialconsensus}
\end{figure}
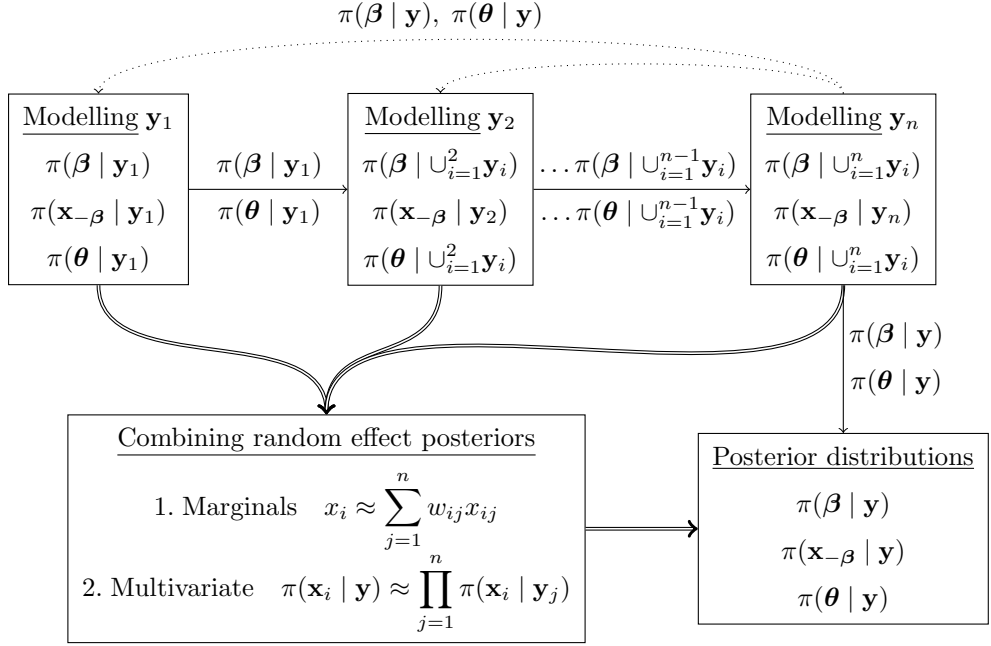
}

\subsection{Sharing latent field components}

Until now we have shown a general framework to implement the sequential consensus in any context. In what follows, we now present how to implement it in the context of integrated models (also known as joint models). The main characteristic of these models is the possibility of sharing random effects by means of setting scaling parameters. For example, a random effect $\mathbf{x}$ can be shared in the linear predictor of another likelihood by scaling it by a parameter $\alpha$ as $\mathbf{x}^*=\alpha \cdot \mathbf{x}$. 

Our proposal to implement sequential consensus with models sharing effects takes into account the following issue with the hyperparameters of the shared random effects. If a random effect, $\mathbf{x}\sim \text{GMRF}(\boldsymbol\mu, \mathbf{Q}(\boldsymbol\theta))$, is scaled in another model as $\alpha\cdot \mathbf{x}$, then the shared random effect is distributed as a $\text{GMRF}(\alpha\cdot\boldsymbol\mu, \alpha^{-2}\cdot\mathbf{Q})$. Since this scaling implies also modifying the precision structure, we must be aware that it is not possible to perform the previously proposed sequential updating for every hyperparameter of that random effect. In fact, if we express the precision matrix of $\mathbf{x}$ as $\mathbf{Q}(\tau, \boldsymbol\theta)=\tau \mathbf{R}(\boldsymbol\theta)$, where $\tau$ is the marginal precision, for $\mathbf{x}^*$ we have $\mathbf{Q}^*=\alpha^{-2}\tau\mathbf{R}(\boldsymbol\theta)=\tau^*\mathbf{R}(\boldsymbol\theta)$. This implies that it is not possible to perform an updating on $\tau$, and so, we can only update the remaining hyperparameters $\boldsymbol\theta$ of the GMRF.

In line with this, we propose two approaches to estimate the sharing parameter $\alpha$: a Gaussian approximation of its posterior distribution and a point estimate of it. The first consists in combining the approximation of the ratio of the marginal distributions for each node. In particular, each of the quotients $\alpha_i = x^*_i/x_i$, where $x^*_i\sim \mathcal{N}(\mu^*_i, \sigma^*_i)$ and $x_i\sim \mathcal{N}(\mu_i, \sigma_i)$, can be considered as a ratio of two Gaussian random variables. Then, following \citep{Hayya_NoteRatioTwoGaussians_1975}, the quotient can be approximated as $\alpha_i \sim \mathcal{N}(\mu^{(\alpha)}_i, \tau^{(\alpha)}_i)$ by means of a second-order Taylor expansion, resulting in a Gaussian distribution with mean and variance respectively 
\begin{equation}
\begin{array}{c}
     \displaystyle \mu^{(\alpha)}_{i} \approx \frac{\mu^*_i}{\mu_i} + \frac{\mu^*_i}{\tau_i\mu_i^3} - \frac{\rho}{\mu^2_i\sqrt{\tau^*_i\tau_i}}\,, \\
     \displaystyle \tau^{(\alpha)}_{i} \approx  \frac{{\mu^*_i}^2}{\tau_i^2\mu_i^4} + \frac{1}{\tau^*_i\mu_i^2} - 2\frac{\rho\mu^*_i}{\mu_i^3\sqrt{\tau^*_i\tau_i}}\,,\\
\end{array}\label{eq:gaussian_approx_alpha}
\end{equation}
where $\rho$ is the correlation between $x^*_i$ and $x_i$. From these approximations of the $\alpha_i$'s, our first proposal to approximate the posterior distribution of the shared effect $\alpha$ is $\pi(\alpha) = \prod_{i}^n \pi(\alpha_i)$. It is worth noting that this approximation tends to be closer to the distribution computed by the integrated model when $\rho = 0$.

Among other tested options, our second approach to estimate the sharing parameter $\alpha$ is to use the empirical median of the following set of values $\{\mu^*_i/\mu_i\,; i = 1,\ldots,n\}$, where $\mu^*_i$ and $\mu_i$ are the mean of $x^*_i$ and $x_i$, respectively. 

Both proposals are based on their relatively good performance in simulated and real scenarios, showing particular accuracy when the shared random effect behaves proportionally across the different models it is shared between. However, when the shared random effects have posterior distributions that are not proportional, according to a scaling parameter, across the different models, the sequential consensus method may yield slightly different results than the integrated model. This discrepancy could suggest that these random effects do not share information about the processes in which they are involved, at least not linearly. 

Once we have a method for estimating the sharing parameter, we can integrate it in the main procedure of the sequential consensus framework. In particular, in order to combine the information of two shared random effects $\mathbf{x}$ and $\mathbf{x}^*$, we have to scale the second effect $\mathbf{x}^*$ by an estimate $\tilde{\alpha}$ of the scaling parameter, that is, $\mathbf{x}^*/\tilde{\alpha}$. This $\tilde{\alpha}$ can be either the point estimate of the second method or the mean of the Gaussian approximation of the first method in Equation (\ref{eq:gaussian_approx_alpha}). The final step is then to perform the consensus between the two random effects $\mathbf{x}^*/\tilde{\alpha}$ and $\mathbf{x}$ as presented in the previous sections.

\subsection{Sequential consensus algorithms}

The sequential process of integrating the information by updating the priors of the fixed effects and hyperparameters and combining the random effects information stored throughout the different sequential inference steps can be synthesised in the Sequential Consensus algorithm (\hyperlink{alg:sequentialconsensus}{SC}, from now on). The algorithm starts with a split dataset or a set of datasets, and allows to perform inference in sequence for those subsets, updating the marginal prior distributions in the step $i$ of the sequence by using the marginal posterior related to each fixed effect and hyperparameter from the previous step $i-1$. In addition, at each step of the sequence, the information related to the random effects is stored. This can be either the marginal posterior distribution, if the weighted marginal averaging approach is applied, or the multivariate posterior distribution for each random effect, if the multivariate Gaussian density product approach is used. 

\begin{algorithm}
\caption{Sequential Consensus (SC)} \label{alg:sequentialconsensus}
\hypertarget{alg:sequentialconsensus}{}
$\mathbf{y} = \{\mathbf{y}_1, ... , \mathbf{y}_n\}$ \Comment{dataset partition} \\
\For{$i = 1$ to $n$}{\vspace{1mm}
    \If{$i = 1$}{\vspace{1mm}
        Inference of the first dataset $\mathbf{y}_1$:\\
        $\pi_{\boldsymbol\beta}[i] \gets \pi(\boldsymbol\beta\mid \mathbf{y}_i) \propto \pi(\mathbf{y}_i\mid \boldsymbol\beta)\pi(\boldsymbol\beta)$ \Comment{Store fixed effect posteriors}\\
        $\pi_{\boldsymbol\theta}[i] \gets \pi(\boldsymbol\theta\mid \mathbf{y}_i)$ \Comment{Store hyperparameter posteriors}\\
        $\pi_{\mathbf{x}}[i] \gets \pi(\mathbf{x}_{-\boldsymbol\beta}\mid \mathbf{y}_i,\boldsymbol\theta)$ \Comment{Store random effect posteriors}\\
    }
    \Else{\vspace{1mm}
        Inference of the $i-th$ dataset, using $\{\pi_{\boldsymbol\beta}[i-1],\pi_{\boldsymbol\theta}[i-1]\}$ as prior distributions:\\
        $\pi_{\boldsymbol\beta}[i] \gets \pi(\boldsymbol\beta\mid \cup_{j=1}^i\mathbf{y}_j)\propto \pi(\boldsymbol\beta\mid \mathbf{y}_i)\pi_{\boldsymbol\beta}[i-1] $ \\
        $\pi_{\boldsymbol\theta}[i] \gets \pi(\boldsymbol\theta\mid \cup_{j=1}^i\mathbf{y}_j)\propto \pi(\boldsymbol\theta\mid \mathbf{y}_i)\pi_{\boldsymbol\theta}[i-1]$ \\
        $\pi_{\mathbf{x}}[i] \gets \pi(\mathbf{x}_{-\boldsymbol\beta}\mid \mathbf{y}_i,\boldsymbol\theta)$ \\
    }
}
\If{Sharing latent field components}{\vspace{1mm}
    Compute an approximation of the scaling parameters $\alpha_i$ using either the Gaussian approximation or the point estimate.
}
\If{Marginal weighted averages}{\vspace{1mm}
    Weighted sum of the random variables $x_{ij}$ from marginal posterior distributions for each $i$ latent field node:\\
    $x_i\mid \mathbf{y} \approx \sum_{j=1}^n w_{ij}x_i[j] \; : \; x_i[j] \sim \pi_{x_i}[j]$ \\
    $$
    \pi(x_i \mid \mathbf{y}) = \mathcal{N}(\mu_i, \tau_i) \; : \; 
    \left\lbrace\begin{array}{l}
        \mu_i=\sum_{j=1}^n w_j \mu_{ij} \\
        \tau_i = \left(\sum_{j=1}^n w^2_j/\tau_{ij}\right)^{-1}  
    \end{array}\right. .
    $$
    
}
\ElseIf{Product of multivariate Gaussian densities}{\vspace{1mm}
    Product of multivariate densities for each random effects structured $\mathcal{X}_i$ stored:\\
    $\pi(\mathbf{x}_i\mid \mathbf{y}) \approx \prod_{j=1}^n \pi_{\mathbf{x}_i}[j]$\\
    $$
    \pi(\mathbf{x}_i \mid \mathbf{y}) = \text{GMRF}(\boldsymbol\mu_i, \mathbf{Q}_i) \; : \; 
    \left\lbrace\begin{array}{l}
        \mathbf{Q}_i = \displaystyle \sum_{j=1}^n \mathbf{Q}_{ij}, \\
        \boldsymbol\mu_{i} = \displaystyle \mathbf{Q}_i^{-1}\sum_{j=1}^n\mathbf{Q}_{ij}\boldsymbol\mu_{ij}. \\ 
    \end{array}\right. .
    $$
}
\end{algorithm}

However, the \hyperlink{alg:sequentialconsensus}{SC} algorithm has a shortcoming that can be very important in certain cases. This is due to the fact that the random effects of the latent field estimated in the first step lack the information on fixed effects and hyperparameters that would be available in the last step of the algorithm. Thus, if a partition of the latent field is also done to reduce the computational burden of each step, similar to the partition of the latent field implemented in \citep{Aritz_BigDM_2023, Aritz_HighDimensionalMultivariate_2023}, we will not be able to perform the consensus procedure between these non-common parts of the latent field. This means that we cannot correct for the lack of information in the estimates of the non-common random effects in the initial steps of the algorithm.

For those situations, we propose to use the following algorithm that avoids that deficiency, the Sequential Consensus for latent field Partitions (\hyperlink{alg:sequentialconsensus2}{SCP}, from now on). This second algorithm allows us to leverage all the information obtained in the first algorithm by performing a second pass through the partition, fixing the posterior distributions of the hyperparameters and re-evaluating the posterior distributions of the fixed effects to avoid using duplicated information in each step. With this new algorithm, we are able to obtain better estimations of the latent field effects by leveraging the computations done in \hyperlink{alg:sequentialconsensus}{SC}. 

The new algorithm starts fixing the posterior distribution of the hyperparameters resulting from the application of algorithm \hyperlink{alg:sequentialconsensus}{SC}. This can be done taking advantage of the own INLA's methodology, as fixing the support points that INLA uses for calculating the marginal posteriors of the latent field is equivalent to fix those posteriors of the hyperparameters. 

The second step in this new proposed algorithm involves the re-evaluation of the fixed effects, as their posterior distribution cannot be fixed. The underlying idea of this step is avoiding the use of duplicated information, and consists of computing the posterior distribution of the fixed effects $\pi(\boldsymbol\beta\mid \mathbf{y}_{-i})$ for the full dataset excluding the data corresponding at each step $i$ of the algorithm that will be used as the prior distribution at that same step. It can be shown that this distribution $\boldsymbol\beta\mid \mathbf{y}_{-i}$ is Gaussian with variance and mean respectively
\begin{equation}
\begin{array}{c}
     \tau^* = \tau_{i-1} + \tau_n - \tau_{i},   \\
     \mu^* = (\tau_{i-1} + \tau_n - \tau_{i})\times(\tau_{i-1} \mu_{i-1} + \tau_n \mu_n - \tau_i\mu_i).
\end{array} \label{eq:prior_fixed_secondpass}
\end{equation}

In order to shown this result in Equation (\ref{eq:prior_fixed_secondpass}), we need to take into account that
\begin{equation}
    \pi(\boldsymbol\beta\mid \mathbf{y}) \propto \prod_{j=1}^{i}\left[\pi(\mathbf{y}_j\mid \boldsymbol\beta) \; \pi(\boldsymbol\beta)\right]\; \times \prod_{j'=i+1}^n \pi(\mathbf{y}_{j'}\mid \boldsymbol\beta),
\end{equation}
where $\prod_{j=1}^{i}\left[\pi(\mathbf{y}_j\mid \boldsymbol\beta)\; \pi(\boldsymbol\beta)\right]$ is also proportional to the posterior $\pi(\boldsymbol\beta \mid \cup_{j=1}^i\mathbf{y}_j)$. 

Then, $\pi(\boldsymbol\beta\mid \mathbf{y}_{-i})$ can be obtained as:
\begin{equation}
\begin{array}{lll}
     \displaystyle \pi(\boldsymbol\beta\mid \mathbf{y}_{-i}) & \propto & \displaystyle \prod_{j=1}^{i-1}\left[\pi(\mathbf{y}_j\mid \boldsymbol\beta)\; \pi(\boldsymbol\beta)\right] \;\times \prod_{j'=i+1}^n \pi(\mathbf{y}_{j'}\mid \boldsymbol\beta) \\
     \displaystyle & \propto & \displaystyle \pi(\boldsymbol\beta\mid \cup_{j=1}^{i-1}\mathbf{y}_j) \; \times \; \frac{\pi(\boldsymbol\beta \mid \mathbf{y})}{\pi(\boldsymbol\beta \mid \cup_{j'=1}^i\mathbf{y}_{j'})}\,.
\end{array}
\end{equation}

Assuming now that $\boldsymbol\beta\mid\mathbf{y}\sim \mathcal{N}(\mu_n, \tau_n)$, $\boldsymbol\beta\mid\cup_{j=1}^i \mathbf{y}_j\sim \mathcal{N}(\mu_i, \tau_i)$ and $\boldsymbol\beta\mid\cup_{j=1}^{i-1} \mathbf{y}_j\sim \mathcal{N}(\mu_{i-1}, \tau_{i-i})$, that is, that these posterior distributions are Gaussian, then the distribution of $\boldsymbol\beta\mid \mathbf{y}_{-i}$ is also Gaussian, $\boldsymbol\beta\mid \mathbf{y}_{-i} \sim \mathcal{N}(\mu^*, \tau^*)$, with the mean and variance in Equation (\ref{eq:prior_fixed_secondpass}).

After re-evaluating the distributions of the fixed parameters and fixing the joint posterior distribution of the hyperparameters, the final step of the \hyperlink{alg:sequentialconsensus2}{SCP} algorithm is to compute the posterior distribution of the latent field for each step. To conclude, note that this algorithm allows a better estimation of the random effects of the latent field, particularly when there are non-common random effects among the partition elements, by taking advantage of all the calculations performed in the \hyperlink{alg:sequentialconsensus}{SC} algorithm.

\begin{algorithm}
\caption{Sequential Consensus for latent field Partitions (SCP)} \label{alg:sequentialconsensus2}
\hypertarget{alg:sequentialconsensus2}{}

$\mathbf{y} = \{\mathbf{y}_1, ... , \mathbf{y}_n\}$ \Comment{dataset partition} \\
\For{$i = 1$ to $n$}{\vspace{1mm}
    Perform the sequential process as in \hyperlink{alg:sequentialconsensus}{SC} Algorithm.
}
\For{$i = 1$ to $n$}{\vspace{1mm}
    Perform the second-pass sequential process. \\
    Use the design support points for the hyperparameters $ \pi(\boldsymbol\theta\mid \mathbf{y})$. \\
    Re-calculating the posterior distribution for the fixed effects:\\
    $$
    \displaystyle \pi(\boldsymbol\beta\mid \mathbf{y}_{-i}) = \pi(\boldsymbol\beta\mid \cup_{j=1}^{i-1}\mathbf{y}_j) \prod_{j'=i+1}^n \pi(\mathbf{y}_{j'}\mid \boldsymbol\beta)
    $$
    Use $\pi(\boldsymbol\beta\mid \mathbf{y}_{-i})$ as the prior distribution for the fixed effects. \\
    Compute and store the latent random effects $\pi(\mathbf{x}_i\mid \mathbf{y}_i)$.
}
\If{Sharing latent field components}{\vspace{1mm}
    Compute an approximation of the scaling parameters $\alpha_i$ using either the Gaussian approximation or the point estimate.
}
\If{Marginal weighted averages}{\vspace{1mm}
    Weighted sum of the random variables $x_{ij}$ from marginal posterior distributions for each $i$ latent field node:\\
    $x_i\mid \mathbf{y} \approx \sum_{j=1}^n w_{ij}x_i[j] \; : \; x_i[j] \sim \pi_{x_i}[j]$ where $\pi(x_i \mid \mathbf{y}) = \mathcal{N}(\mu_i, \tau_i)$.
}
\ElseIf{Product of multivariate Gaussian densities}{\vspace{1mm}
    Product of multivariate densities for each random effects structured $\mathcal{X}_i$ stored:\\
    $\pi(\mathbf{x}_i\mid \mathbf{y}) \approx \prod_{j=1}^n \pi_{\mathbf{x}_i}[j]$ where $\pi(\mathbf{x}_i \mid \mathbf{y}) = \text{GMRF}(\boldsymbol\mu_i, \mathbf{Q}_i)$.
}
\end{algorithm}

\section{Examples}

In this section, we illustrate the application of the sequential consensus inference methodology for analysing different data sets. We first present a simulated example using a spatio-temporal model with two different sampling designs (stratified random sampling and preferential sampling) to show how sequential consensus can be used to combine data from different sampling processes. We also present two examples of real data: one from fisheries science, where the integrated model is complex and involves the integration of different sources of information (biomass, abundance and presence-absence), and another involving a large amount of temperature data. The latter example allows us to illustrate the use of sequential consensus for handling large databases.

\subsection{Simulated example: Integrating data from different sampling designs}

In this example, we simulate a scenario where the abundance of a species is sampled through two surveys with different sampling designs: (i) a stratified random design and (ii) a positive preferential sampling design. Such scenarios are common in fisheries science, and integrated models have been used to analyse this particular case \citep{Alglave_CombiningIndPref_2022}.

In this first case, we have performed the simulation using a simple model within a square survey region $\mathcal{D}=\{(0,10)\times(0,10)\}$ for $10$ temporal realisations or nodes:
\begin{equation}
\begin{array}{c}
     y_i \mid \eta_i, \boldsymbol\theta \sim \text{Gamma}(y_i \mid \eta_i, \tau)\,, \\
     \log(\mu_i) = \beta_0 + \beta_1 \cdot x_i + f_t(z_i) + u_i\,,
\end{array}
\end{equation}
where $\mu_i$ is the mean of the Gamma distribution $\log(\mu_i)=\eta_i$, $\beta_0$ is an intercept, $\beta_1$ is a linear coefficient related to a covariate $x_i$ (which can be a variable such as temperature or bathymetry), $f_t(z_i)$ denotes a second-order random walk effect for the temporal component (assumed to be evenly spaced in months or years), and $u_i$ represents a common spatially structured effect across different time nodes. $\boldsymbol\theta$ is the vector of hyperparameters, including the precision for the Gamma likelihood $\tau$, and the hyperparameters for the latent field random effects. As discussed in the sub-section on spatio-temporal models, the structure of this model allows us to analyse a persistent spatial distribution with random intensity changes over time.

\begin{figure}
    \centering
    \includegraphics[width = \linewidth]{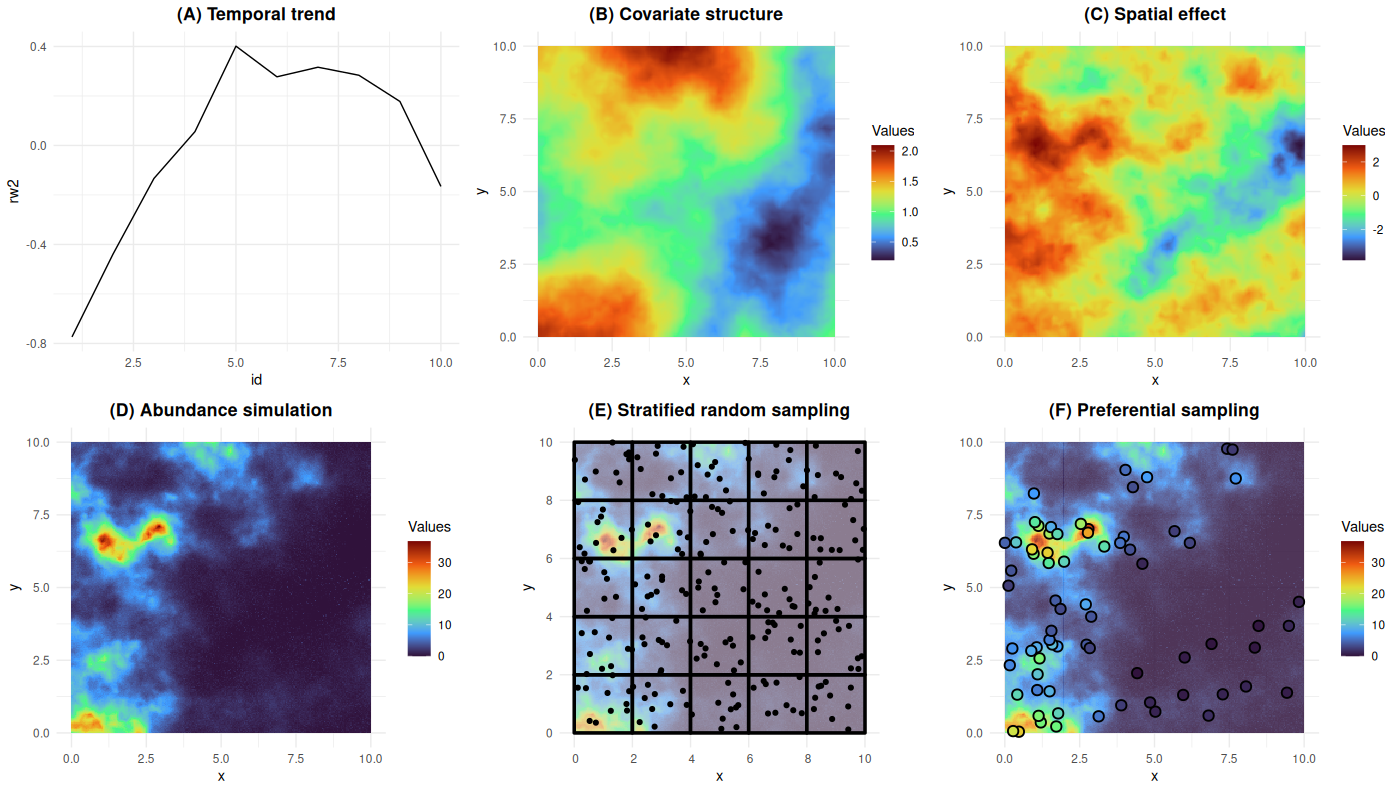}
    \caption{Simulated example. Simulated temporal, spatial, and spatially structured covariate effects on top panels. Below, the response variable (abundance) for the first temporal node and simulated sampling based on stratified random sampling and preferential sampling.}
    \label{fig:simulation_samplings_designs}
\end{figure}

The simulation of the samples involves defining a grid for the stratified random sampling design and implementing a LGCP for the preferential sampling. For the stratified random sampling, we have created a grid of $25$ cells within the square survey region, as shown in Figure \ref{fig:simulation_samplings_designs}. In each cell, 10 samples are simulated for each of the $10$ time nodes. For the preferential sampling we have simulated the pattern from the following LGCP model
\begin{equation}
\begin{array}{c}
     y(s_i) \mid \lambda(s_i) \sim \text{LGCP}(\lambda(s_i)) \,,\\
     \log(\lambda(s_i)) = \beta^s_0 + \alpha \cdot f_t(z_i) + u_i   \,,
\end{array}
\label{eq:lgcp_simulation}
\end{equation}
where $\beta^s_0$ is a global intercept that allows us to control the amount of samples simulated over the entire spatio-temporal setting. In the LGCP linear predictor, the spatial term is shared with the gamma linear predictor without scaling, while the temporal component $f_t(z_i)$ has the same values as those the used in the abundance simulation but is scaled by an $\alpha$ parameter. The $\boldsymbol\theta^*$ represents the vector of the hyperparameters for the LGCP linear predictor.

In equation (\ref{eq:lgcp_simulation}) we set the value of $\beta_0^*$ by considering that the two sampling designs will have approximately the same number of samples over the spatial and temporal implementation. We then set the expected sample size equal to the number of samples simulated in the completely random stratified design $\Lambda = 25\cdot10\cdot10$; $10$ samples per cell, $25$ cells per time node and $10$ time nodes. So we have $\beta_0^*$ calculated as
\begin{equation}
\begin{array}{rcl}
     \beta_0^* & = & \displaystyle \log\left(\Lambda\right) - \log\left(\iint \exp(\alpha\cdot f_t(z_i) + u_i)\text{d}s\text{d}t \right), \\
     & \approx & \displaystyle \log\left(\Lambda\right) - \log\left(\sum_{i=1}^n \exp(\alpha\cdot f_t(z_i) + u_i)\Delta_i \right),
\end{array}
\end{equation}
where $\Delta_i$ is the area associated with each spatial element in each temporal node.

In this example, we have approximated two different integrated models, one that includes shared components without scaling across them, and another one that fits scaled predictors. The first integrated model with equal latent fields looks like this:
\begin{equation}
\begin{array}{rcl}
     y^{srs}_i \mid \eta^{srs}_i, \tau_g &\sim& \text{Gamma}(y^{srs}_i \mid \eta^{srs}_i, \tau_g),  \\
     \log(\mu_i) &=& \beta_0 + \beta_1 \cdot x_i + f_t(z_i) + u_i, \\
     y^{ps}_j \mid \eta^{ps}_j, \tau_g &\sim& \text{Gamma}(y^{ps}_j \mid \eta^{ps}_j, \tau_g), \\
     \log(\mu_j) &=& \beta_0 + \beta_1 \cdot x_j + f_t(z_j) + u_j, \\
     y(s_j) \mid \lambda(s_j) &\sim& \text{LGCP}(\lambda(s_j)), \\
     \log(\lambda(s_j)) &=& \beta^s_0 + f^*_t(z_j) + u_j
\end{array}
\end{equation}
where $y^{srs}_i$ are the response variable values from the completely random stratified sampling and $y^{ps}_j$ are the values coming from the preferential sampling. 

The second model includes a scaled temporal effect for the marks and the point process $\log(\lambda_j) = \beta^s_0 + \alpha\cdot f_t(z_j) + u_j$. In any case, since it is scaled by a parameter $\alpha$, we can also make a point estimate of this parameter or obtain an approximation of its distribution, as discussed in the section on the sequential consensus algorithm.

\begin{figure}
    \centering
    \includegraphics[width=\linewidth]{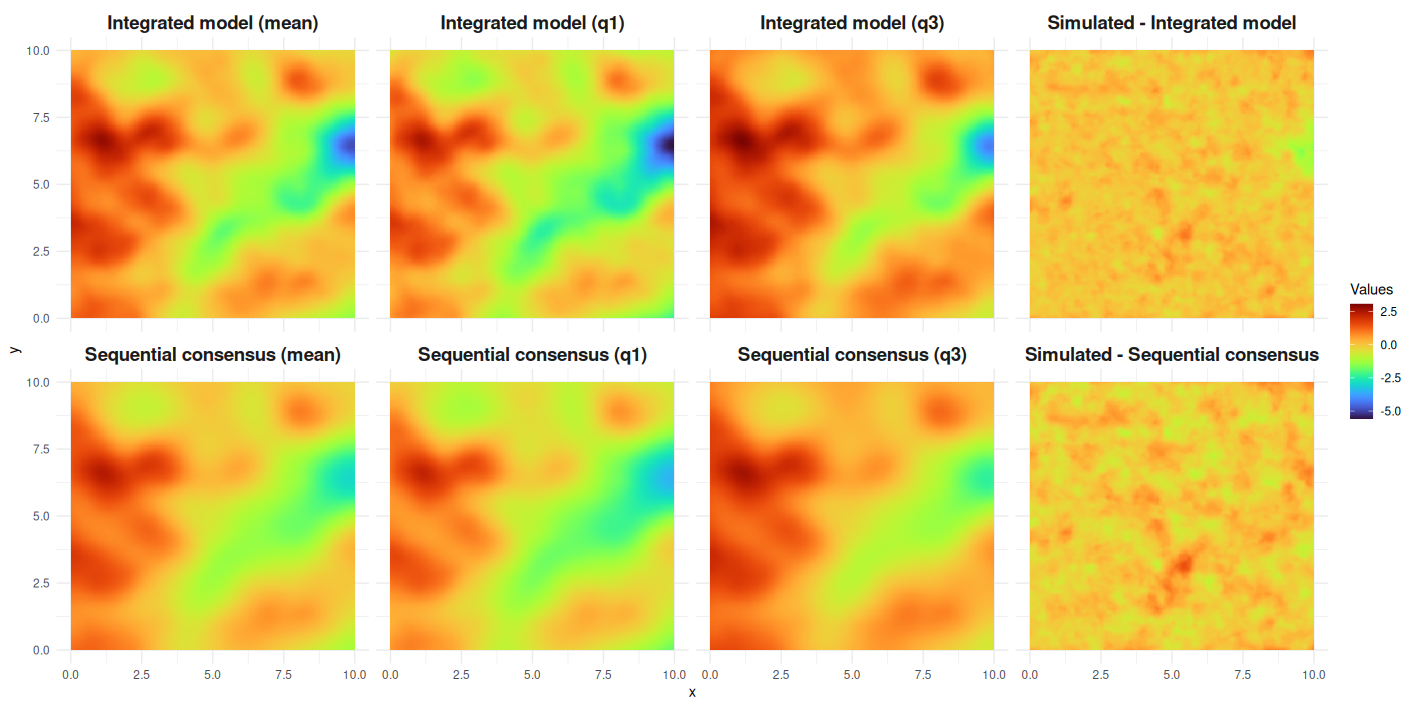}
    \caption{Simulated example. Spatial effect results obtained by the integrated model and those obtained by the consensus method. The plots show the means and quantiles at $0.025$ and $0.975$ of the posterior distributions. In addition, the last two vertical plots on the right show the difference between the simulated spatial field and the mean of the posterior distribution obtained by both the integrated model and the sequential consensus.}
    \label{fig:spatial_resultsEx1}
\end{figure}

\begin{figure}
    \centering
    \includegraphics[width=\linewidth]{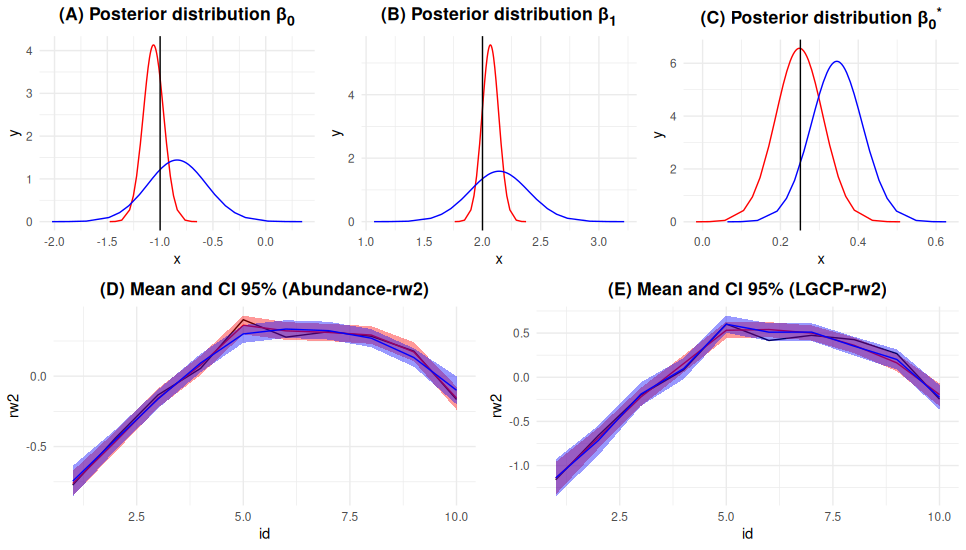}
    \caption{Simulated example. Posterior distributions of the fixed effects are displayed, along with the mean and $95\%$ credible intervals for the posterior distribution of the temporal trends. Distributions estimated from the integrated model are shown in red, while those estimated from the sequential consensus are shown in blue.}
    \label{fig:posterior_distributionsEx1}
\end{figure}

\begin{figure}
    \centering
    \includegraphics[width = \linewidth]{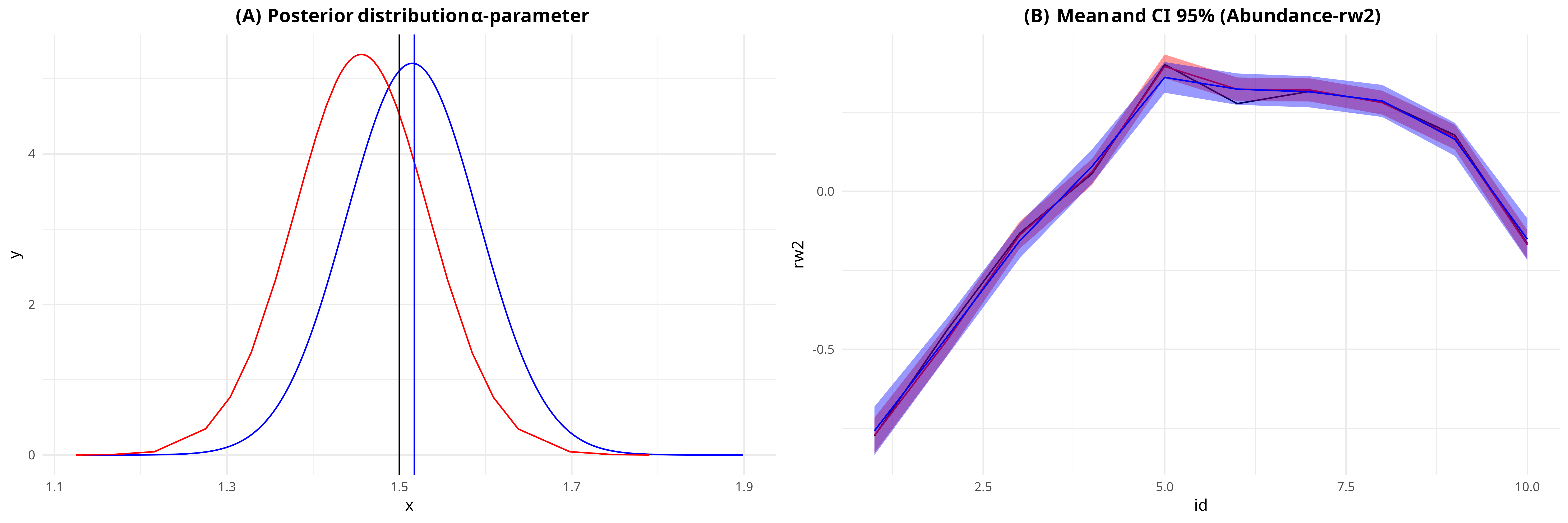}
    \caption{Simulated example. Posterior distributions of the time effect (right panel) and the alpha scaling parameter (left panel). In red the posterior distribution obtained using the integrated model, while in blue the approximation obtained using the distributions given in the sequential consensus. Vertical lines refer to the sequential consensus point estimate (blue) and the true value (black).}
    \label{fig:alpha_rw2_v2}
\end{figure}

The results show that the sequential consensus procedure yields very similar results to the integrated model with equal latent fields (Figures \ref{fig:spatial_resultsEx1} and \ref{fig:posterior_distributionsEx1}) and the integrated model with scaled latent fields (Figure \ref{fig:alpha_rw2_v2}). However, when estimating the hyperparameters of the model, the sequential consensus procedure shows differences with respect to the results obtained using a integrated model (Table \ref{tab:table_hyperparEx1}). This discrepancy arises because the assumption used to approximate the posterior distribution of the hyperparameters is not precise enough to replicate the results of the integrated model.

The results for the second integrated model, including scaled joint predictors, differ primarily in the estimation of the temporal effect, as shown in Figure \ref{fig:alpha_rw2_v2}. In addition, the approximate distribution of the parameter $\alpha$ for both the sequential consensus approach and the integrated model is shown in Figure \ref{fig:alpha_rw2_v2}, along with the point estimate for the sequential consensus. 

\begin{table}
    \centering
    \begin{tabular}{l|r|r|r|r|r|r}
    \hline
    \multicolumn{7}{c}{Integrated model hyperparameters} \\ \hline
      & mean & sd. & 0.025 quant. & 0.5 quant. & 0.975 quant. & mode\\ \hline
    $\tau$ & 3.16 & 0.06 & 3.04 & 3.16 & 3.29 & 3.16\\ \hline
    $\tau^A_{rw2}$ & 35.60 & 18.22 & 9.74 & 32.27 & 79.07 & 24.96\\ \hline
    $\theta_1$ & 0.40 & 0.18 & 0.08 & 0.39 & 0.78 & 0.34\\ \hline
    $\theta_2$ & 0.33 & 0.16 & 0.05 & 0.32 & 0.68 & 0.28\\ \hline
    $\tau^{LGCP}_{rw2}$ & 40.89 & 27.77 & 6.85 & 34.19 & 110.55 & 19.88\\ \hline \hline
    \multicolumn{7}{c}{Sequential consensus hyperparameters} \\ \hline
      & mean & sd. & 0.025 quant. & 0.5 quant. & 0.975 quant. & mode\\ \hline
    $\tau$ & 1.87 & 0.01 & 1.85 & 1.87 & 1.89 & 1.87\\ \hline
    $\tau^A_{rw2}$ & 38.73 & 0.21 & 38.32 & 38.73 & 39.14 & 38.73\\ \hline
    $\theta_1$ & 0.38 & 0.01 & 0.37 & 0.38 & 0.39 & 0.38\\ \hline
    $\theta_2$ & 0.22 & 0.01 & 0.21 & 0.22 & 0.23 & 0.22\\ \hline
    $\tau^{LGCP}_{rw2}$ & 1.01 & 0.12 & 1.05 & 1.31 & 1.52 & 1.34\\ \hline 
    \end{tabular}
    \caption{Simulated example. Summary of posterior hyperparameters for the integrated model and the sequential consensus procedure.}
    \label{tab:table_hyperparEx1}
\end{table}

On the computational side, the integrated model took 2.97 minutes to complete, while the sequential consensus process took 0.99 minutes. These computations were performed on a laptop with 16 GB of RAM and a 2.3 GHz AMD Ryzen 7 3700u processor. As shown in the following examples, computational efficiency improves significantly as model complexity or dataset size increases.

\subsection{Analysing hake distribution in the bay of Biscay}

\begin{figure}
    \centering
    \includegraphics[width = 0.7\linewidth]{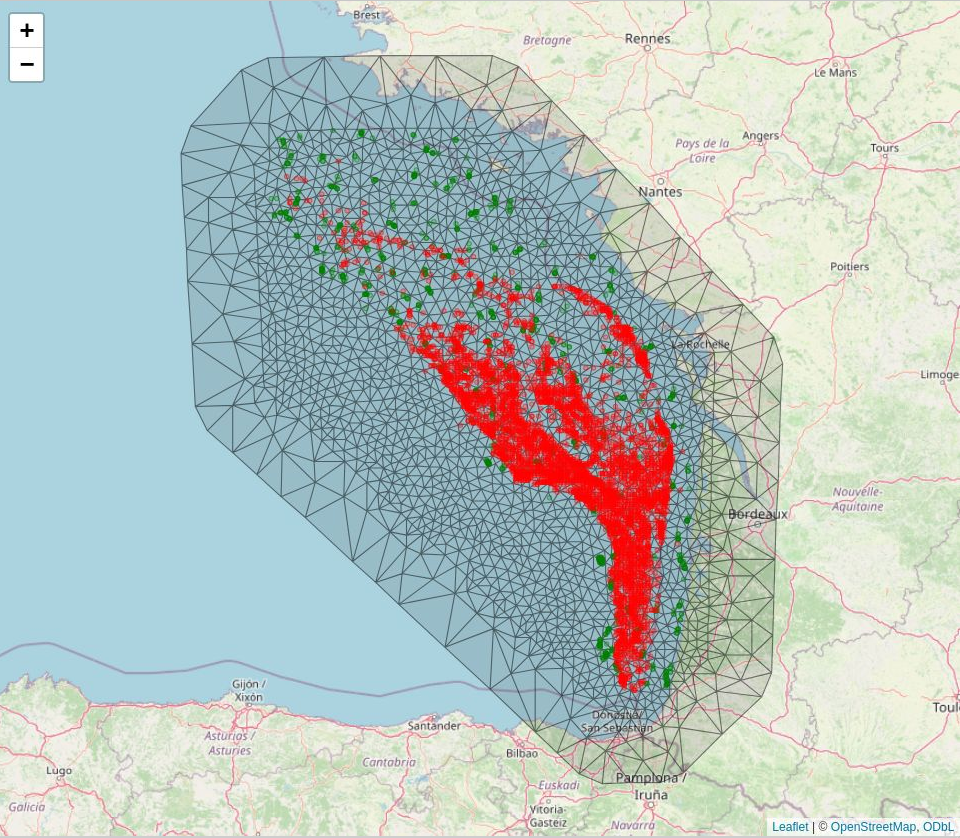}
    \caption{Hake example. Data from non-preferential random sampling (green dots) and preferential sampling (red dots) obtained in surveys during 2003 and 2021 along the Bay of Biscay.}
    \label{fig:RealData_Bay_Biscay}
\end{figure}

This case study shows how different sources of information can be combined. In particular, with the ultimate aim of describing the spatial distribution of hake, we combine data from the EVHOE trawl survey and two commercial fishing fleets, sampled by on-board observers, for the years 2003 to 2021 in the Bay of Biscay (Figure \ref{fig:RealData_Bay_Biscay}). 

The scientific survey collected discrete abundance data, while one commercial fishing fleet targeting hake collected continuous biomass data and the other commercial fleet recorded presence-absence data. The commercial fleet targeting hake carried out a preferential exploration of the sea in order to maximise the hake biomass catch. This implies a preferential sampling process, so the integrated model that takes into account all this information, including the one under preferential sampling, can be expressed as follows:
\begin{equation}
\begin{array}{c}
    y_{1i} \mid \eta_{1i} \sim \text{Po}(\lambda_i)    \\
    \log(\lambda_i) = \beta_{10} + f_{1d}(z_{di}) + f_{1y}(z_{yi}) + \alpha_{s1} \cdot u_i, \\
    y_{2j} \mid \eta_{2j}, \tau \sim \text{Gamma}(y_i \mid \eta_{2j},\tau), \\
    \log(\mu_j) = \beta_{20} + f_{2d}(z_{dj}) + f_{2y}(z_{yj}) + u_j, \\
    y_{3j} \mid \eta_{3j} \sim \text{Ber}(\pi_j), \\
    \text{logit}(\pi_j) = \beta_{30} + \alpha_{d3} \cdot f_{2d}(z_{dj}) + f_{3y}(z_{yj}) + \alpha_{s3}\cdot u_j, \\
    y(s_{j}) \mid \lambda(s_j) \sim \text{LGCP}(\lambda(s_j)), \\
    \log(\lambda(s_j)) = \beta_{40} + \alpha_{d4} \cdot f_{2d}(z_{dj}) + u^*_j, 
\end{array}
\end{equation}
where the count data from the scientific survey ($\mathbf{y}_1$) is Poisson distributed, the preferentially sampled biomass ($\mathbf{y}_2$) follows a Gamma distribution and the presence/absence ($\mathbf{y}_{3}$) follows a Bernoulli distribution. The point pattern of the preferential sampling pattern ($\mathbf{s}$) is modelled using a LGCP. The $\boldsymbol\beta$ parameters are intercepts associated with each of the response variables, $\boldsymbol\alpha$ components represent scaling parameters for the shared effects, $f_d$ represents a structured random effect related to the depth covariate (a second-order random walk for $f_{1d}$ and a one-dimensional SPDE for $f_{2d}$), and $f_y$ is a first-order random walk for years. Finally, $\mathbf{u}$ is a separable type III spatio-temporal effect \citep{KnorrHeld_SpaceTime_2000}, with a precision matrix derived from a two-dimensional SPDE with Mátern covariances \citep{Lindgren_ExplicitLinkSPDE_2011} for the spatial part and an iid precision matrix for the temporal part.

In implementing the sequential consensus approach, we can consider different ways of splitting the integrated model, e.g. we could consider splitting it into as many parts as there are likelihoods, or we could separate all parts except the gamma distribution of the preferential data and the LGCP. In this case, we decompose the integrated model according to the second proposal and obtain three partitions: a preferential model with biomass data, a Poisson model for counts and a Bernoulli model for presence-absences.

\begin{figure}
    \centering
    \includegraphics[width=\linewidth]{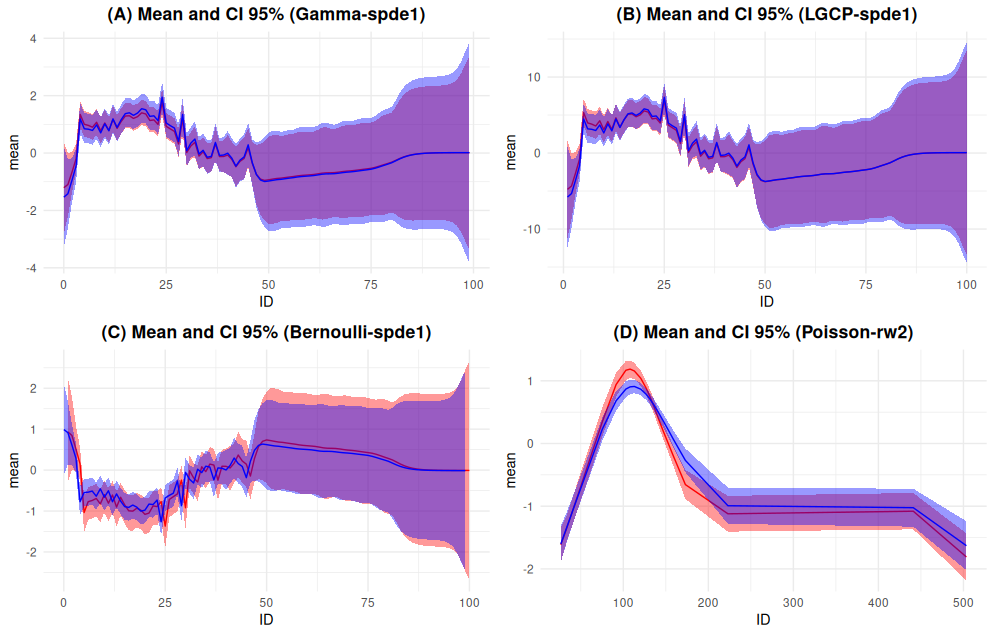}
    \caption{Hake example. Mean and $95\%$ credible intervals for the posterior distribution of depth. Distributions estimated from the integrated model are shown in red, while those estimated from the sequential consensus are shown in blue.}
    \label{fig:depth_posterior_Hake}
\end{figure}

\begin{figure}
    \centering
    \includegraphics[width=\linewidth]{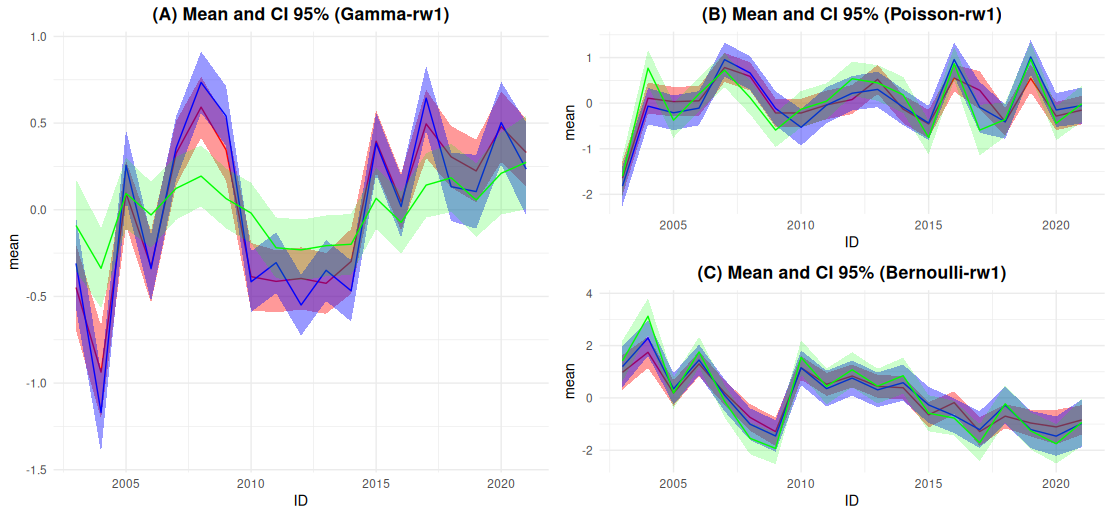}
    \caption{Hake example. Mean and $95\%$ credible intervals for the posterior distribution for the time effect. Distributions estimated from the integrated model are shown in red, while those estimated from the sequential consensus with the correction from the multivariate consensus are shown in blue and those estimates with the correction from the marginal consensus are shown in green.}
    \label{fig:Posterior_year_Hake}
\end{figure}

\begin{figure}
    \centering
    \includegraphics[width=\linewidth]{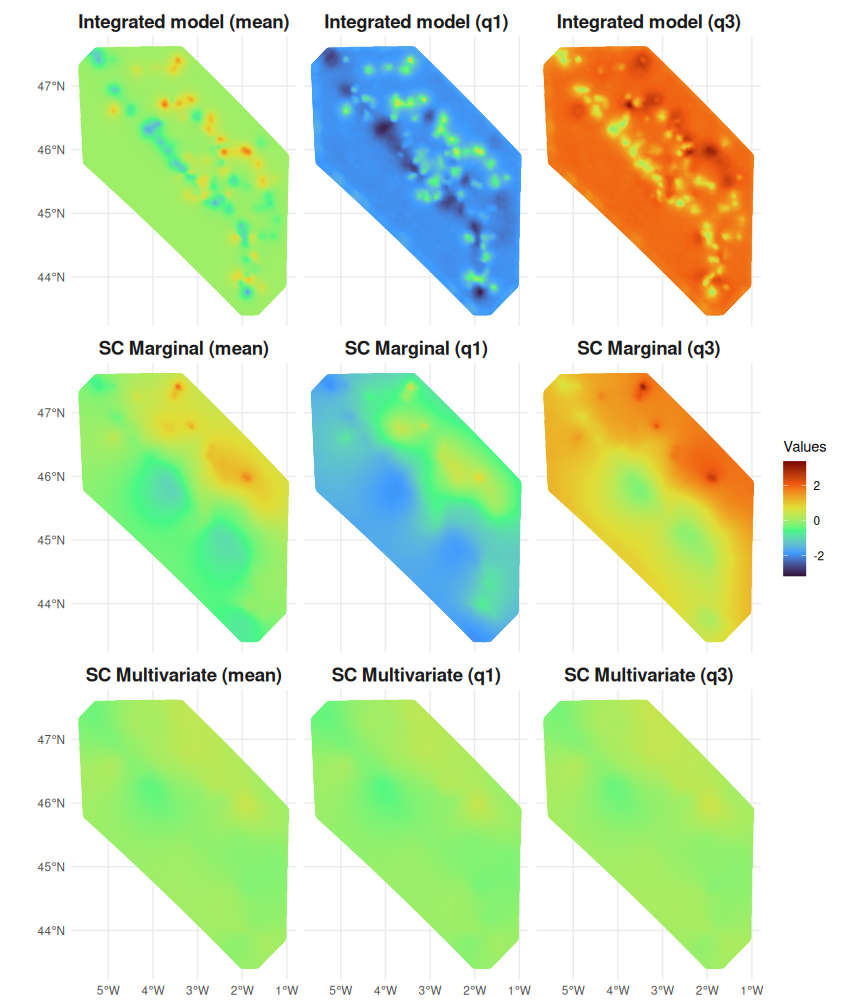}
    \caption{Hake example. Spatial-temporal effect for the first temporal node of the shared spatio-temporal component between the Gamma, Poisson and Bernoulli models. The plots display the mean values, as well as the quantiles at $0.025$ and $0.975$, of the posterior distributions, comparing also the marginal and the multivariate consensus.}
    \label{fig:Spatiotemporal_posterior_Hake}
\end{figure}

\begin{figure}
    \centering
    \includegraphics[width=\linewidth]{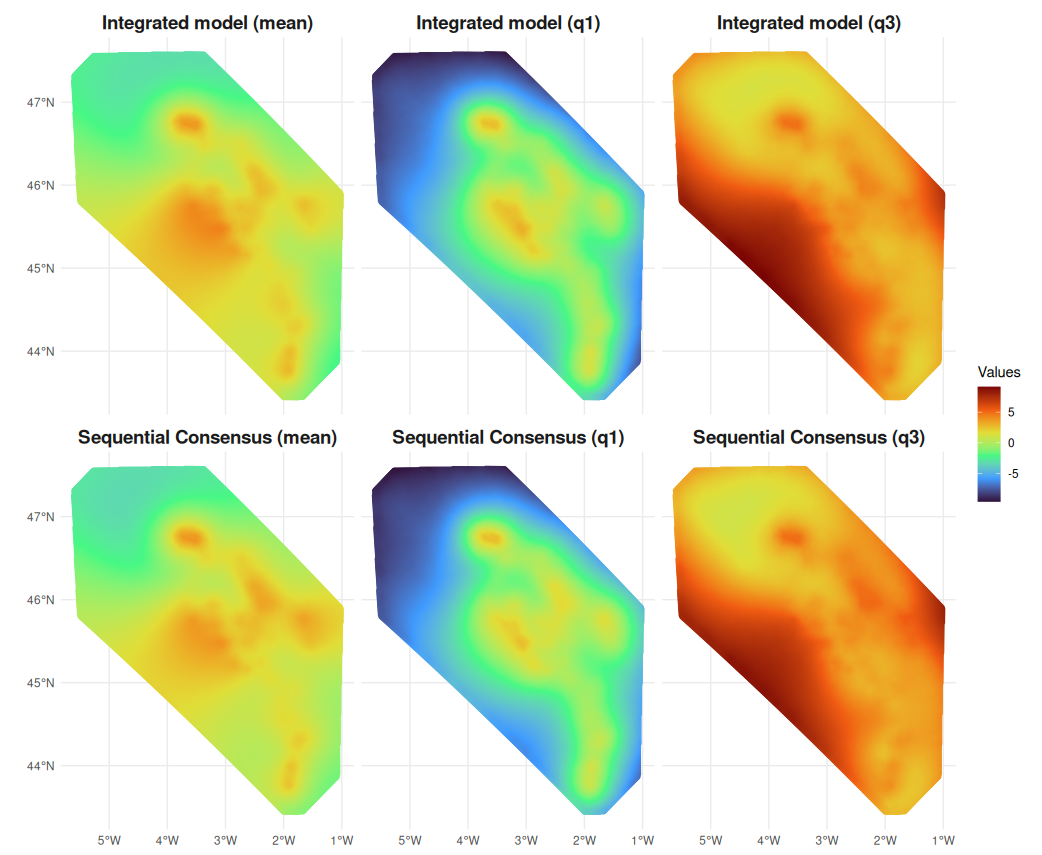}
    \caption{Hake example. Spatial-temporal effect for the first temporal node of the spatio-temporal component for the LGCP. The plots display the mean values, as well as the quantiles at $0.025$ and $0.975$, of the posterior distributions.}
    \label{fig:Spatiotemporal_posterior_LGCP_Hake}
\end{figure}

The results for the latent field are shown in the following figures. In Figure \ref{fig:Posterior_fixed_Hake} we show the posterior distribution for the fixed effects, where the main discrepancy between the integrated model and the sequential consensus comes from the intercept for the Bernoulli response variable. In Figure \ref{fig:depth_posterior_Hake} we show the mean and the $95\%$ credible interval (CI) for the effect of depth. Figure \ref{fig:Posterior_year_Hake} shows the mean and $95\%$ CI for the temporal effects. To compute the pure temporal structure, a post-consensus correction is applied to the spatio-temporal component, since the difference in spatial aggregation for each temporal node after consensus induces a change in the pure temporal component. This means that this correction is applied depending on the two types of consensus proposed, marginal or multivariate. Finally, in Figure \ref{fig:Spatiotemporal_posterior_Hake} and \ref{fig:Spatiotemporal_posterior_LGCP_Hake} we show the posterior distribution for the spatio-temporal component for the shared spatio-temporal term and the spatio-temporal effect for the LGCP, respectively. 

In terms of computational time, the integrated model takes $62.12$ minutes, while the sequential consensus approach performed by \hyperlink{alg:sequentialconsensus}{SC} takes $13.81$ minutes, with all computations performed on a server with $63$ cores and $157$ GB of RAM.

\begin{figure}
    \centering
    \includegraphics[width=\linewidth]{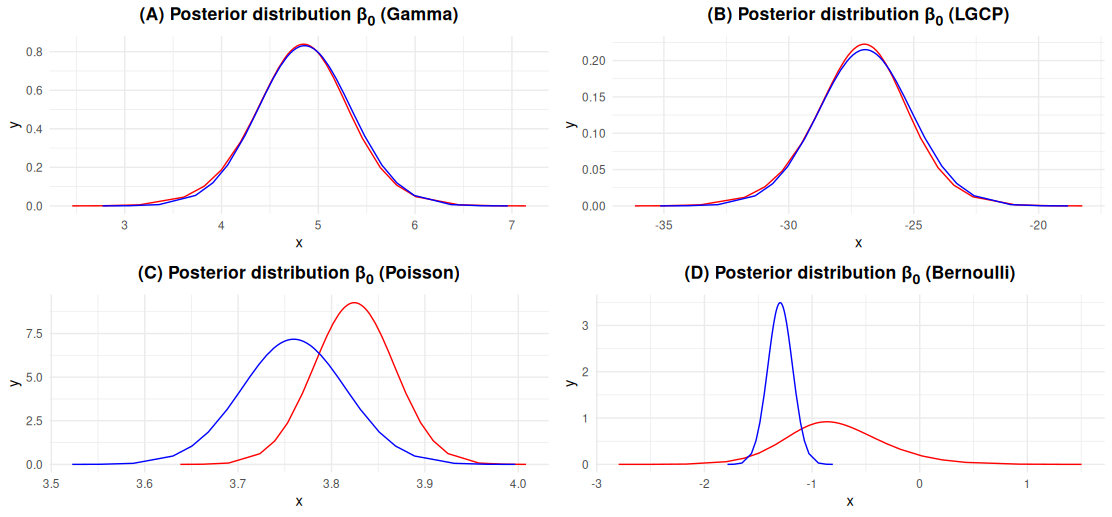}
    \caption{Hake example. Fixed effects. In red the posterior distributions from the integrated model and in blue the posterior distribution from the sequential consensus approach.}
    \label{fig:Posterior_fixed_Hake}
\end{figure}

\subsection{Spatio-temporal temperature modelling}

\begin{figure}
    \centering
    \includegraphics[width = \linewidth]{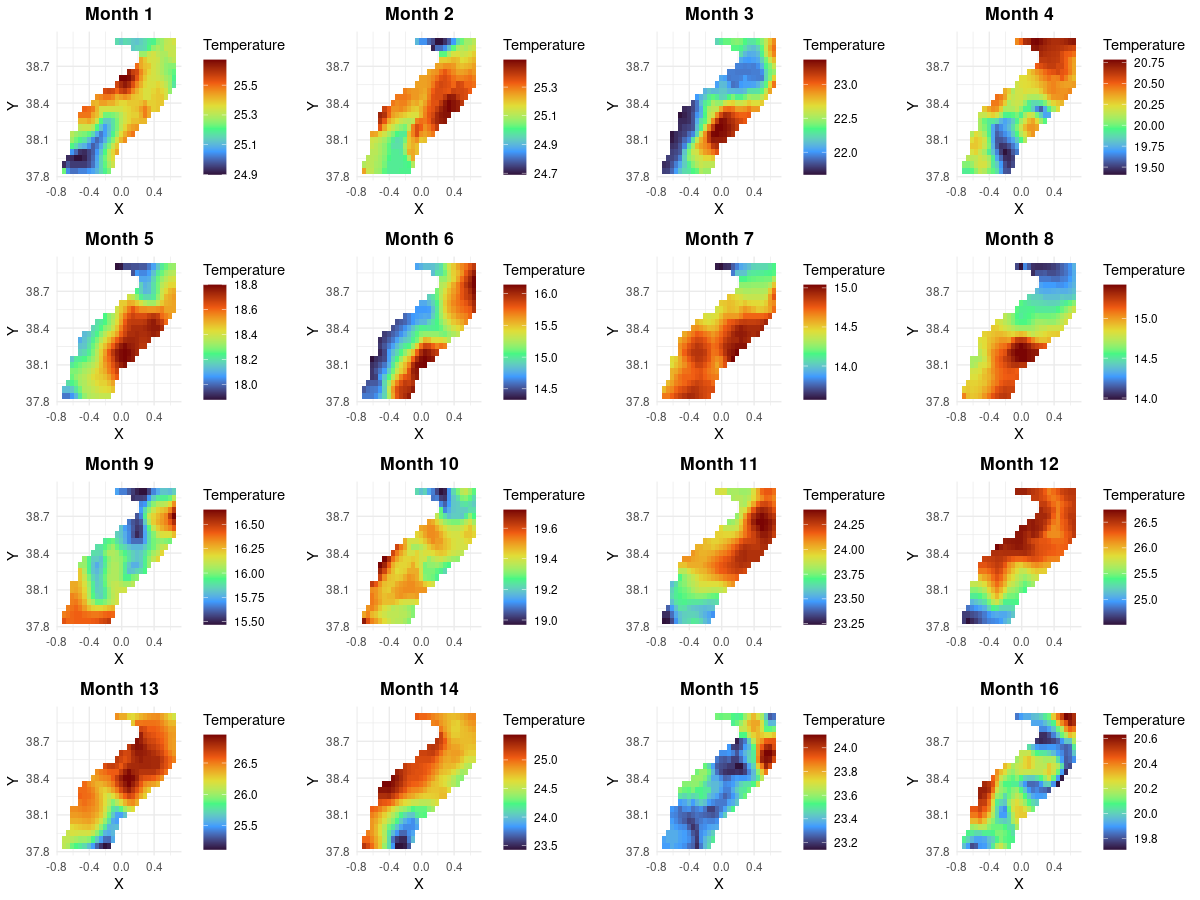}
    \caption{Temperature Big Data example. Representation of the first 16 months for temperature values.}
    \label{fig:geolocated_dataExTemp}
\end{figure}

Our final example illustrates the use of sequential consensus to deal with large databases. In particular, we analyse temperature measurements at $308$ geolocated locations collected over $480$ months in the coastal area of Alicante, Spain, as shown in the Figure \ref{fig:geolocated_dataExTemp}. It can be seen that the temperature values have a spatial structure, but this varies over the months without there appearing to be any real pattern that repeats systematically.

The aim of this example is to show how the sequential consensus approach can be used when the partitioning of the data breaks the latent field to reduce its computational burden. In fact, the results of the complete model are compared with the results obtained with the two sequential consensus algorithms (\hyperlink{alg:sequentialconsensus}{SC} and \hyperlink{alg:sequentialconsensus2}{SCP}) applied to the partitioned data sets. The comparison is based on the posterior distributions of the latent field nodes, the posterior of the hyperparameters, and the computational cost of the different modelling approaches.

In this case we use a separable spatio-temporal interaction model $\mathbf{Q}_{st}=\mathbf{Q}_s\otimes\mathbf{Q}_t$. The precision matrix of the spatial structure is constructed according to a two-dimensional SPDE effect with Matérn's covariance function \citep{Lindgren_ExplicitLinkSPDE_2011}, while the precision matrix associated with the temporal structure is defined according to an autoregressive effect of order 1 structure. The spatio-temporal model can be written as:
\begin{equation}
\begin{array}{c}
     y_i \mid \eta_i, \tau \sim \mathcal{N}(y_i \mid \eta_i, \tau), \\
     \mu_i = \beta_0 + u_i, \\
     u_i \sim \text{GRMF}(\mathbf{0}, \mathbf{Q}_{st}),
\end{array}
\end{equation}
where $\mu_i$ is the mean of the normal distribution $\mu_i = \eta_i$, $\tau$ is the precision of the normal distribution, $\beta_0$ is a global intercept and $u_i$ is the spatio-temporal interacting effect with $\mathbf{Q}_{st}(\theta_1, \theta_2, \rho)$ precision matrix, being $\theta_1$ and $\theta_2$ the reparametrization of the spatial range and the standard marginal deviation of the spatial effect, and $\rho$ the autocorrelation hyperparameter.

The model for the complete data set does not run without the process being terminated on a server with 63 cores and 157 GB of RAM. Therefore, we use a subset of the first 120 months to allow for model comparison. This subset is used to run a model on the full data and a sequential consensus approach using the two algorithms. To perform the sequential consensus process, the subset of the $120$ months is divided into $6$ groups, each consisting of the data associated with $20$ consecutive time nodes.

\begin{figure}
    \centering
    \includegraphics[width = \linewidth]{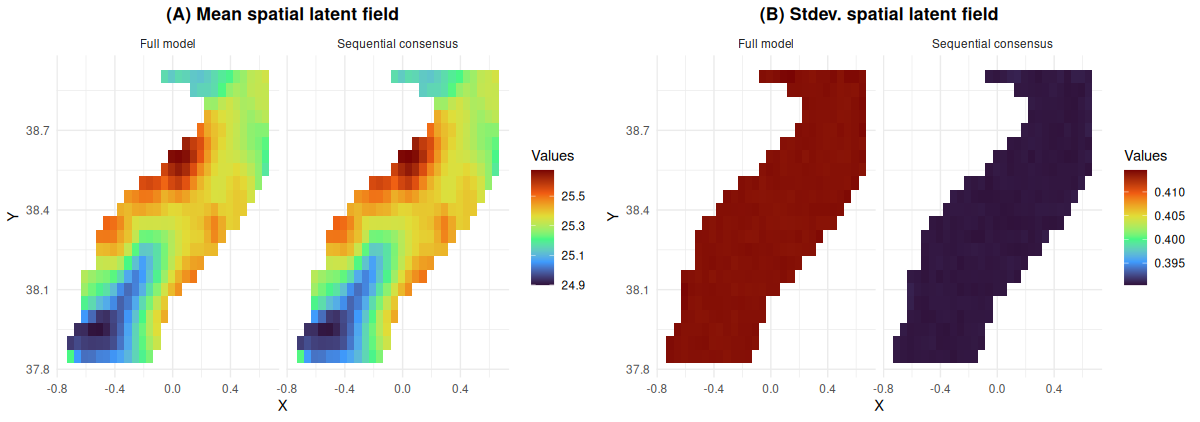}
    \caption{Temperature Big Data example. Mean and standard deviation of the posterior distribution for the spatio-temporal effect for the first month (temporal node), using the \protect\hyperlink{alg:sequentialconsensus2}{SCP} Algorithm.}
    \label{fig:mean_stdev_latentfield_Temp}
\end{figure}

\begin{figure}
    \centering
    \includegraphics[width = \linewidth]{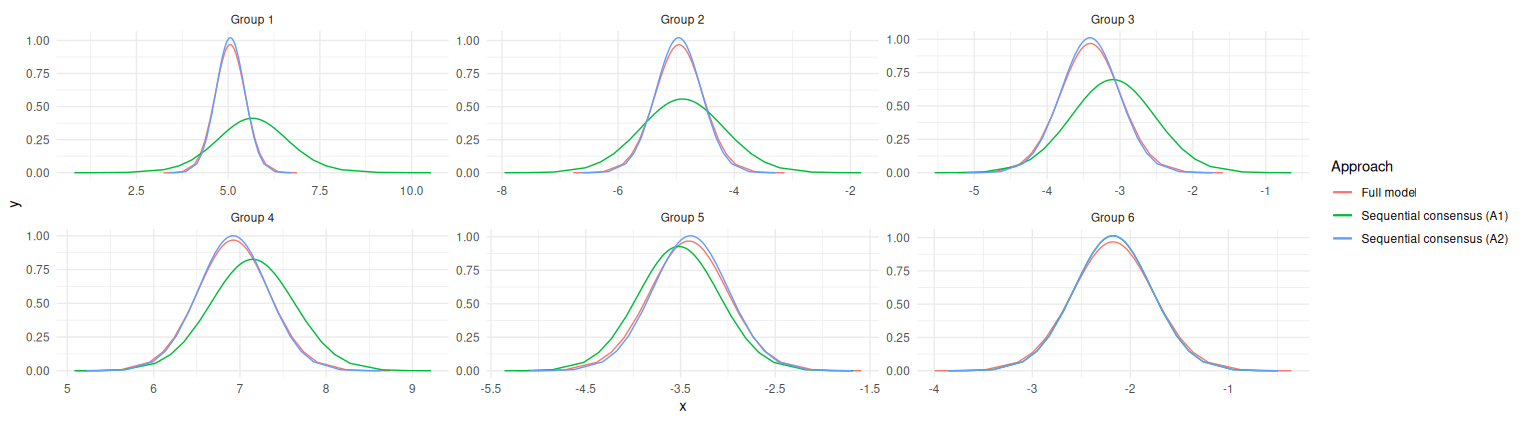}
    \caption{Temperature Big Data example. The marginal posterior distribution of an arbitrary node within each group, into which the data and the latent field have been divided for the sequential consensus approach, is compared according to the corresponding posterior distribution for the model with the complete data and latent field.}
    \label{fig:posterior_latentfield_ExTemperature}
\end{figure}

\begin{figure}
    \centering
    \includegraphics[width = \linewidth]{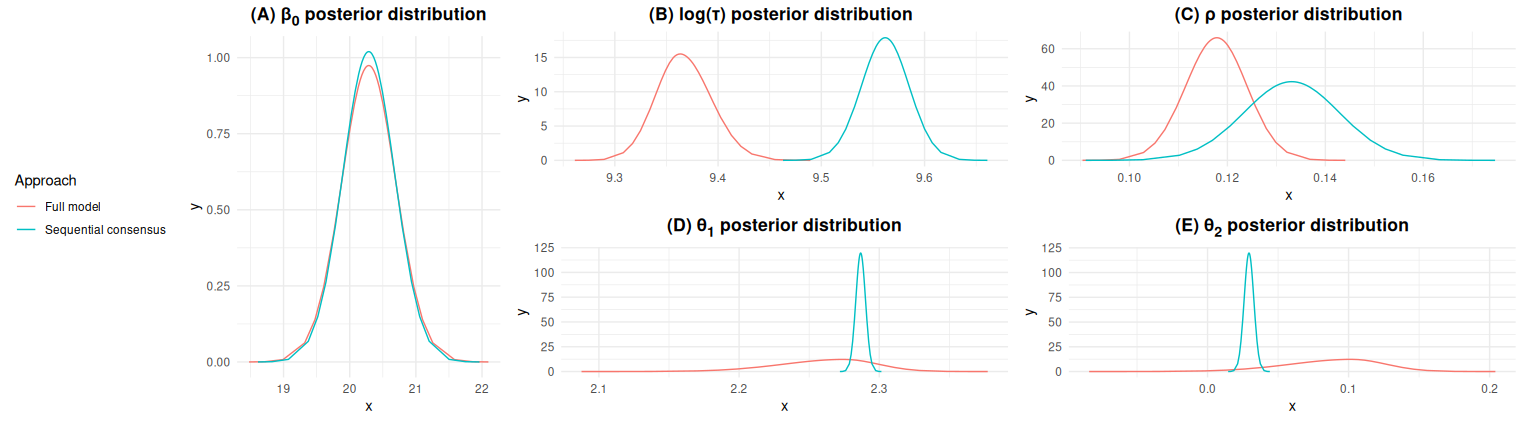}
    \caption{Temperature Big Data example. Posterior distribution for the fixed effect and hyperparameters.}
    \label{fig:posterior_fiexhyperpar_ExTemperature}
\end{figure}

In Figure \ref{fig:mean_stdev_latentfield_Temp}, we can compare the mean and standard deviation of the latent field for the spatio-temporal effect of the first month (as an example) between the complete model and the sequential consensus approach performed according to the Algorithm \hyperlink{alg:sequentialconsensus}{SC}. It can be observed that both results are very similar in terms of mean and standard deviation, with the standard deviation obtained by the sequential consensus Algorithm \hyperlink{alg:sequentialconsensus2}{SCP} being slightly lower. In Figure \ref{fig:posterior_latentfield_ExTemperature} we compare an arbitrary node from each group into which the sequential consensus data have been split with the corresponding node from the complete model. It can be seen that the \hyperlink{alg:sequentialconsensus2}{SCP} Algorithm produces the same results as the complete model, while the \hyperlink{alg:sequentialconsensus}{SC} Algorithm shows a progressive approximation to the complete model as it progresses through the sequence, obtaining the same results as the \hyperlink{alg:sequentialconsensus}{SCP} for the last sequence. Figure \ref{fig:posterior_fiexhyperpar_ExTemperature} presents the posterior distributions of the fixed effect and hyperparameters, showing discrepancies in the latter between the complete model and the sequential consensus approach. 

The computational cost for the complete model is $47.13$ minutes, while for the sequential consensus using the \hyperlink{alg:sequentialconsensus}{SC} Algorithm takes $6.57$ minutes, and using the \hyperlink{alg:sequentialconsensus2}{SCP} Algorithm $11.02$ ($6.57 + 4.45$) minutes. These computations are performed on a server with the previously mentioned specifications: $63$ cores and $157$ GB of RAM. More importantly, as mentioned above, the model does not run with the full set of $480$ time nodes, but with the implementation of the sequential consensus approach, we are able to analyse it in $25.48$ minutes and $37.55$ minutes using the \hyperlink{alg:sequentialconsensus}{SC} Algorithm and the \hyperlink{alg:sequentialconsensus2}{SCP} Algorithm, respectively.

\section{Conclusions}

The landscape of ecological research has been revolutionised by the influx of diverse and rich datasets. The integration of these datasets has great potential to refine our understanding and management of ecological systems \citep{Fletcher_CombiningData_2019}. Integrated models provide a formal framework for combining different types of data and sampling methods. However, as data complexity and volume increase, the computational requirements of integrated models escalate proportionally, posing a significant challenge to practical implementation. In this line, this study proposes a computationally efficient approximation using sequential inference and exploiting the methodological foundations of INLA \citep{Rue_INLA_2009, VanNiekert_NewINLA_2023}, together with the Latent Gaussian Model (LGM) structure.

The case studies included in this study focus on complex spatio-temporal modelling scenarios to showcase the potential of sequential modelling in addressing complex real-world phenomena. Spatio-temporal models are quintessential example of computational burden, where the complexity of these models increases as multiple datasets are integrated, each contributing additional layers of information and nuance. This computational improvement is evident in all three examples, but is particularly significant in the second (from 61.02 minutes to 13.81 minutes) and third examples (from 47.13 minutes to 6.57 minutes using the \hyperlink{alg:sequentialconsensus}{SC} and an additional 4.45 minutes for the \hyperlink{alg:sequentialconsensus2}{SCP}). In these cases, the complexity of the model and/or the size of the data may lead to computationally expensive or even infeasible analyses. In particular, the third example highlights situations where this problem is particularly pronounced.  

The sequential consensus has been approached through different approximations for the fixed components of the latent field and the hyperparameters. For the fixed effects we define their prior distributions according to the posteriors of the sequence's previous inferential process. For the random effects of the latent field we have applied two consensus process approaches, one based on the marginals for each random effect and another based on the multivariate distribution of the random effects. Both consensus proposals lead to similar results, but the multivariate one takes correlations into account and therefore produces better results when correlations have a high impact. After the completion of the sequencing, the accuracy and precision of estimates may not be very good in the first steps of the sequence given that not all data has been used at that stage. Therefore, two algorithms have been proposed, a simpler one applicable when our interest is in the final step of the sequence (e.g. last years abundance estimates) and a refined version that performs a second pass over the partitions, improving the estimation of all random effect estimations. This is clearly demonstrated in the results of Figure \ref{fig:posterior_latentfield_ExTemperature}, where it can be observed that the estimation of the marginal distributions of the random effects was worse in the initial steps of the sequence when Algorithm \hyperlink{alg:sequentialconsensus}{SC} was applied. In contrast, this deficiency is corrected with Algorithm \hyperlink{alg:sequentialconsensus2}{SCP}, with both algorithms converging to the same result by the final step of the sequential process.

In general, both algorithms gave good estimates of the latent field, including both fixed and random effects. However, the \hyperlink{alg:sequentialconsensus}{SC} algorithm gives worse estimates of the random effects in the early steps of the sequence if these are not compensated by estimates from later steps as in the proposed \hyperlink{alg:sequentialconsensus2}{SCP}. However, in terms of reproducing the posterior distributions of the hyperparameters, neither algorithm succeeds in reproducing those obtained by the integrated model. This is because the updating of the marginals requires the assumption of no correlation between the hyperparameters, which is generally not true, although it is sufficiently accurate to allow the correct estimation of the latent field. In addition, the approach used for scaling parameters between shared effects can be problematic if the different posterior distributions in the sequence have non-proportional values. Therefore, when using the sequential consensus approach to mimic integrated models with a large number of likelihoods, it is more effective to recompose the problem using multiple joint models rather than a single large joint model or many single likelihood models.

Note that this sequential consensus allows us to perform a sequential inference updating the priors of the inferential step $i$ with the posteriors obtained in the inferential step $i-1$, and combining the information from the random effects by a consensus strategy. However, since we are updating the marginal distributions, it is expected that for the hyperparameters, this may not be the best approximation compared to updating their joint distribution, due to the high correlation they may exhibit in the posterior joint distribution.

Future developments should improve the updating of hyperparameter information throughout the sequence, together with the implementation of a method to estimate these scaling parameters within the sequential modelling itself. This would minimise the impact of model splitting and allow the integrated model to be reconstructed from the simpler models without loss of fidelity.

\section*{Acknowledgments}

DC, MF and ALQ thank support by the grant PID2022-136455NB-I00, funded by Ministerio de Ciencia, Innovación y Universidades of Spain (MCIN/AEI/10.13039/501100011033/FEDER, UE) and the European Regional Development Fund. DC also acknowledges Grant CIAICO/2022/165 funded by Generalitat Valenciana. 

\bibliography{bibliography}


\begin{thebibliography}{61}
\ifx \bisbn   \undefined \def \bisbn  #1{ISBN #1}\fi
\ifx \binits  \undefined \def \binits#1{#1}\fi
\ifx \bauthor  \undefined \def \bauthor#1{#1}\fi
\ifx \batitle  \undefined \def \batitle#1{#1}\fi
\ifx \bjtitle  \undefined \def \bjtitle#1{#1}\fi
\ifx \bvolume  \undefined \def \bvolume#1{\textbf{#1}}\fi
\ifx \byear  \undefined \def \byear#1{#1}\fi
\ifx \bissue  \undefined \def \bissue#1{#1}\fi
\ifx \bfpage  \undefined \def \bfpage#1{#1}\fi
\ifx \blpage  \undefined \def \blpage #1{#1}\fi
\ifx \burl  \undefined \def \burl#1{\textsf{#1}}\fi
\ifx \doiurl  \undefined \def \doiurl#1{\url{https://doi.org/#1}}\fi
\ifx \betal  \undefined \def \betal{\textit{et al.}}\fi
\ifx \binstitute  \undefined \def \binstitute#1{#1}\fi
\ifx \binstitutionaled  \undefined \def \binstitutionaled#1{#1}\fi
\ifx \bctitle  \undefined \def \bctitle#1{#1}\fi
\ifx \beditor  \undefined \def \beditor#1{#1}\fi
\ifx \bpublisher  \undefined \def \bpublisher#1{#1}\fi
\ifx \bbtitle  \undefined \def \bbtitle#1{#1}\fi
\ifx \bedition  \undefined \def \bedition#1{#1}\fi
\ifx \bseriesno  \undefined \def \bseriesno#1{#1}\fi
\ifx \blocation  \undefined \def \blocation#1{#1}\fi
\ifx \bsertitle  \undefined \def \bsertitle#1{#1}\fi
\ifx \bsnm \undefined \def \bsnm#1{#1}\fi
\ifx \bsuffix \undefined \def \bsuffix#1{#1}\fi
\ifx \bparticle \undefined \def \bparticle#1{#1}\fi
\ifx \barticle \undefined \def \barticle#1{#1}\fi
\bibcommenthead
\ifx \bconfdate \undefined \def \bconfdate #1{#1}\fi
\ifx \botherref \undefined \def \botherref #1{#1}\fi
\ifx \url \undefined \def \url#1{\textsf{#1}}\fi
\ifx \bchapter \undefined \def \bchapter#1{#1}\fi
\ifx \bbook \undefined \def \bbook#1{#1}\fi
\ifx \bcomment \undefined \def \bcomment#1{#1}\fi
\ifx \oauthor \undefined \def \oauthor#1{#1}\fi
\ifx \citeauthoryear \undefined \def \citeauthoryear#1{#1}\fi
\ifx \endbibitem  \undefined \def \endbibitem {}\fi
\ifx \bconflocation  \undefined \def \bconflocation#1{#1}\fi
\ifx \arxivurl  \undefined \def \arxivurl#1{\textsf{#1}}\fi
\csname PreBibitemsHook\endcsname

\bibitem[\protect\citeauthoryear{Fletcher~Jr.
  et~al.}{2019}]{Fletcher_CombiningData_2019}
\begin{barticle}
\bauthor{\bsnm{Fletcher~Jr.}, \binits{R.J.}},
\bauthor{\bsnm{Hefley}, \binits{T.J.}},
\bauthor{\bsnm{Robertson}, \binits{E.P.}},
\bauthor{\bsnm{Zuckerberg}, \binits{B.}},
\bauthor{\bsnm{McCleery}, \binits{R.A.}},
\bauthor{\bsnm{Dorazio}, \binits{R.M.}}:
\batitle{{A practical guide for combining data to model species
  distributions}}.
\bjtitle{Ecology}
\bvolume{100}(\bissue{6}),
\bfpage{02710}
(\byear{2019})
\doiurl{10.1002/ecy.2710}
{\href{https://arxiv.org/abs/https://esajournals.onlinelibrary.wiley.com/doi/pdf/10.1002/ecy.2710}{{https://esajournals.onlinelibrary.wiley.com/doi/pdf/10.1002/ecy.2710}}}
\end{barticle}
\endbibitem

\bibitem[\protect\citeauthoryear{Rufener et~al.}{2021}]{Rufener_Bridging_2021}
\begin{barticle}
\bauthor{\bsnm{Rufener}, \binits{M.-C.}},
\bauthor{\bsnm{Kristensen}, \binits{K.}},
\bauthor{\bsnm{Nielsen}, \binits{J.R.}},
\bauthor{\bsnm{Bastardie}, \binits{F.}}:
\batitle{{Bridging the gap between commercial fisheries and survey data to
  model the spatiotemporal dynamics of marine species}}.
\bjtitle{Ecological Applications}
\bvolume{31}(\bissue{8}),
\bfpage{02453}
(\byear{2021})
\doiurl{10.1002/eap.2453}
{\href{https://arxiv.org/abs/https://esajournals.onlinelibrary.wiley.com/doi/pdf/10.1002/eap.2453}{{https://esajournals.onlinelibrary.wiley.com/doi/pdf/10.1002/eap.2453}}}
\end{barticle}
\endbibitem

\bibitem[\protect\citeauthoryear{Alglave
  et~al.}{2022}]{Alglave_CombiningIndPref_2022}
\begin{barticle}
\bauthor{\bsnm{Alglave}, \binits{B.}},
\bauthor{\bsnm{Rivot}, \binits{E.}},
\bauthor{\bsnm{Etienne}, \binits{M.-P.}},
\bauthor{\bsnm{Woillez}, \binits{M.}},
\bauthor{\bsnm{Thorson}, \binits{J.T.}},
\bauthor{\bsnm{Vermard}, \binits{Y.}}:
\batitle{{Combining scientific survey and commercial catch data to map fish
  distribution}}.
\bjtitle{ICES Journal of Marine Science}
\bvolume{79}(\bissue{4}),
\bfpage{1133}--\blpage{1149}
(\byear{2022})
\doiurl{10.1093/icesjms/fsac032}
{\href{https://arxiv.org/abs/https://academic.oup.com/icesjms/article-pdf/79/4/1133/43783429/fsac032.pdf}{{https://academic.oup.com/icesjms/article-pdf/79/4/1133/43783429/fsac032.pdf}}}
\end{barticle}
\endbibitem

\bibitem[\protect\citeauthoryear{Paradinas et~al.}{2023}]{Paradinas_ISDM_2023}
\begin{barticle}
\bauthor{\bsnm{Paradinas}, \binits{I.}},
\bauthor{\bsnm{Illian}, \binits{J.B.}},
\bauthor{\bsnm{Alonso-Fernändez}, \binits{A.}},
\bauthor{\bsnm{Pennino}, \binits{M.G.}},
\bauthor{\bsnm{Smout}, \binits{S.}}:
\batitle{{Combining fishery data through integrated species distribution
  models}}.
\bjtitle{ICES Journal of Marine Science}
\bvolume{80}(\bissue{10}),
\bfpage{2579}--\blpage{2590}
(\byear{2023})
\doiurl{10.1093/icesjms/fsad069}
\end{barticle}
\endbibitem

\bibitem[\protect\citeauthoryear{Isaac
  et~al.}{2020}]{Isaac_DataIntegration_2020}
\begin{barticle}
\bauthor{\bsnm{Isaac}, \binits{N.J.}},
\bauthor{\bsnm{Jarzyna}, \binits{M.A.}},
\bauthor{\bsnm{Keil}, \binits{P.}},
\bauthor{\bsnm{Dambly}, \binits{L.I.}},
\bauthor{\bsnm{Boersch-Supan}, \binits{P.H.}},
\bauthor{\bsnm{Browning}, \binits{E.}},
\bauthor{\bsnm{Freeman}, \binits{S.N.}},
\bauthor{\bsnm{Golding}, \binits{N.}},
\bauthor{\bsnm{Guillera-Arroita}, \binits{G.}},
\bauthor{\bsnm{Henrys}, \binits{P.A.}}, \betal:
\batitle{{Data integration for large-scale models of species distributions}}.
\bjtitle{Trends in ecology \& evolution}
\bvolume{35}(\bissue{1}),
\bfpage{56}--\blpage{67}
(\byear{2020})
\doiurl{10.1016/j.tree.2019.08.006}
\end{barticle}
\endbibitem

\bibitem[\protect\citeauthoryear{Fithian et~al.}{2015}]{Fithian_Bias_2015}
\begin{barticle}
\bauthor{\bsnm{Fithian}, \binits{W.}},
\bauthor{\bsnm{Elith}, \binits{J.}},
\bauthor{\bsnm{Hastie}, \binits{T.}},
\bauthor{\bsnm{Keith}, \binits{D.A.}}:
\batitle{Bias correction in species distribution models: pooling survey and
  collection data for multiple species}.
\bjtitle{Methods in Ecology and Evolution}
\bvolume{6}(\bissue{4}),
\bfpage{424}--\blpage{438}
(\byear{2015})
\doiurl{10.1111/2041-210X.12242}
\end{barticle}
\endbibitem

\bibitem[\protect\citeauthoryear{Ara{\'u}jo and
  New}{2007}]{Araujo_Ensemble_2007}
\begin{barticle}
\bauthor{\bsnm{Ara{\'u}jo}, \binits{M.B.}},
\bauthor{\bsnm{New}, \binits{M.}}:
\batitle{Ensemble forecasting of species distributions}.
\bjtitle{Trends in ecology \& evolution}
\bvolume{22}(\bissue{1}),
\bfpage{42}--\blpage{47}
(\byear{2007})
\doiurl{10.1016/j.tree.2006.09.010}
\end{barticle}
\endbibitem

\bibitem[\protect\citeauthoryear{Nisbet
  et~al.}{2018}]{Nisbet_HandbookStatisticalAnalysis_2018}
\begin{bbook}
\bauthor{\bsnm{Nisbet}, \binits{R.}},
\bauthor{\bsnm{Miner}, \binits{G.D.}},
\bauthor{\bsnm{Yale}, \binits{K.}}:
\bbtitle{{H}andbook of {S}tatistical {A}nalysis and {D}ata {M}ining
  {A}pplications}.
\bpublisher{Academic Press},
\blocation{Boston}
(\byear{2018}).
\doiurl{10.1016/B978-0-12-416632-5.09980-1} .
\burl{https://www.sciencedirect.com/science/article/pii/B9780124166325099801}
\end{bbook}
\endbibitem

\bibitem[\protect\citeauthoryear{Fletcher and
  Fortin}{2019}]{Fletcher_SpatialEcology_2019}
\begin{bbook}
\bauthor{\bsnm{Fletcher}, \binits{R.}},
\bauthor{\bsnm{Fortin}, \binits{M.-J.}}:
\bbtitle{Spatial Ecology and Conservation Modeling: Applications with R}.
\bpublisher{Springer},
\blocation{Cham}
(\byear{2019}).
\doiurl{10.1007/978-3-030-01989-1}
\end{bbook}
\endbibitem

\bibitem[\protect\citeauthoryear{Mevin B.~Hooten and
  Brost}{2021}]{Hooten_RecursiveBayesianInference_2021}
\begin{barticle}
\bauthor{\bsnm{Mevin B.~Hooten}, \binits{D.S.J.}},
\bauthor{\bsnm{Brost}, \binits{B.M.}}:
\batitle{{Making Recursive Bayesian Inference Accessible}}.
\bjtitle{The American Statistician}
\bvolume{75}(\bissue{2}),
\bfpage{185}--\blpage{194}
(\byear{2021})
\doiurl{10.1080/00031305.2019.1665584}
{\href{https://arxiv.org/abs/https://doi.org/10.1080/00031305.2019.1665584}{{https://doi.org/10.1080/00031305.2019.1665584}}}
\end{barticle}
\endbibitem

\bibitem[\protect\citeauthoryear{Doucet
  et~al.}{2010}]{Doucet_SequentialMonteCarlo_2001}
\begin{bbook}
\bauthor{\bsnm{Doucet}, \binits{A.}},
\bauthor{\bsnm{Freitas}, \binits{N.}},
\bauthor{\bsnm{Gordon}, \binits{N.}},
\bauthor{\bsnm{Smith}, \binits{A.}}:
\bbtitle{Sequential Monte Carlo Methods in Practice},
\bedition{Softcover reprint of hardcover 1st ed. 2001} edn.
\bsertitle{Information Science and Statistics}.
\bpublisher{Springer},
\blocation{New York}
(\byear{2010})
\end{bbook}
\endbibitem

\bibitem[\protect\citeauthoryear{Scott
  et~al.}{2016}]{Scott_ConsensusMonteCarlo_2016}
\begin{barticle}
\bauthor{\bsnm{Scott}, \binits{S.L.}},
\bauthor{\bsnm{Blocker}, \binits{A.W.}},
\bauthor{\bsnm{Bonassi}, \binits{F.V.}},
\bauthor{\bsnm{Chipman}, \binits{H.A.}},
\bauthor{\bsnm{George}, \binits{E.I.}},
\bauthor{\bsnm{McCulloch}, \binits{R.E.}}:
\batitle{{Bayes and big data: the consensus Monte Carlo algorithm}}.
\bjtitle{International Journal of Management Science and Engineering
  Management}
\bvolume{11}(\bissue{2}),
\bfpage{78}--\blpage{88}
(\byear{2016})
\doiurl{10.1080/17509653.2016.1142191}
{\href{https://arxiv.org/abs/https://doi.org/10.1080/17509653.2016.1142191}{{https://doi.org/10.1080/17509653.2016.1142191}}}
\end{barticle}
\endbibitem

\bibitem[\protect\citeauthoryear{Scott}{2017}]{Scott_ConsensusDistributedComputation_2017}
\begin{barticle}
\bauthor{\bsnm{Scott}, \binits{S.L.}}:
\batitle{{Comparing consensus Monte Carlo strategies for distributed Bayesian
  computation}}.
\bjtitle{Brazilian Journal of Probability and Statistics}
\bvolume{31}(\bissue{4}),
\bfpage{668}--\blpage{685}
(\byear{2017})
\doiurl{10.1214/17-BJPS365}
\end{barticle}
\endbibitem

\bibitem[\protect\citeauthoryear{K{\"o}rding and
  Wolpert}{2004}]{Kording_BayesianIntegration_2004}
\begin{barticle}
\bauthor{\bsnm{K{\"o}rding}, \binits{K.P.}},
\bauthor{\bsnm{Wolpert}, \binits{D.M.}}:
\batitle{{Bayesian integration in sensorimotor learning}}.
\bjtitle{Nature}
\bvolume{427}(\bissue{6971}),
\bfpage{244}--\blpage{247}
(\byear{2004})
\doiurl{10.1038/nature02169}
\end{barticle}
\endbibitem

\bibitem[\protect\citeauthoryear{Zigler
  et~al.}{2013}]{Zigler_ModelFeedbackBiometric_2013}
\begin{barticle}
\bauthor{\bsnm{Zigler}, \binits{C.M.}},
\bauthor{\bsnm{Watts}, \binits{K.}},
\bauthor{\bsnm{Yeh}, \binits{R.W.}},
\bauthor{\bsnm{Wang}, \binits{Y.}},
\bauthor{\bsnm{Coull}, \binits{B.A.}},
\bauthor{\bsnm{Dominici}, \binits{F.}}:
\batitle{Model feedback in bayesian propensity score estimation}.
\bjtitle{Biometrics}
\bvolume{69}(\bissue{1}),
\bfpage{263}--\blpage{273}
(\byear{2013})
\doiurl{10.1111/j.1541-0420.2012.01830.x}
\end{barticle}
\endbibitem

\bibitem[\protect\citeauthoryear{Nguyen
  et~al.}{2022}]{Ngunyen_BayesianAbstentionFeedbacks_2022}
\begin{barticle}
\bauthor{\bsnm{Nguyen}, \binits{C.V.}},
\bauthor{\bsnm{Ho}, \binits{L.S.T.}},
\bauthor{\bsnm{Xu}, \binits{H.}},
\bauthor{\bsnm{Dinh}, \binits{V.}},
\bauthor{\bsnm{Nguyen}, \binits{B.T.}}:
\batitle{{Bayesian active learning with abstention feedbacks}}.
\bjtitle{Neurocomputing}
\bvolume{471},
\bfpage{242}--\blpage{250}
(\byear{2022})
\doiurl{10.1016/j.neucom.2021.11.027}
\end{barticle}
\endbibitem

\bibitem[\protect\citeauthoryear{Brakhane
  et~al.}{2012}]{Brakhane_BayesianFeedbackQuantumPhysics_2012}
\begin{barticle}
\bauthor{\bsnm{Brakhane}, \binits{S.}},
\bauthor{\bsnm{Alt}, \binits{W.}},
\bauthor{\bsnm{Kampschulte}, \binits{T.}},
\bauthor{\bsnm{Martinez-Dorantes}, \binits{M.}},
\bauthor{\bsnm{Reimann}, \binits{R.}},
\bauthor{\bsnm{Yoon}, \binits{S.}},
\bauthor{\bsnm{Widera}, \binits{A.}},
\bauthor{\bsnm{Meschede}, \binits{D.}}:
\batitle{{Bayesian Feedback Control of a Two-Atom Spin-State in an Atom-Cavity
  System}}.
\bjtitle{Phys. Rev. Lett.}
\bvolume{109},
\bfpage{173601}
(\byear{2012})
\doiurl{10.1103/PhysRevLett.109.173601}
\end{barticle}
\endbibitem

\bibitem[\protect\citeauthoryear{Figueira
  et~al.}{2024}]{Figueira_BayesianFeedback_2023}
\begin{botherref}
\oauthor{\bsnm{Figueira}, \binits{M.}},
\oauthor{\bsnm{Barber}, \binits{X.}},
\oauthor{\bsnm{Conesa}, \binits{D.}},
\oauthor{\bsnm{López-Quílez}, \binits{A.}},
\oauthor{\bsnm{Martínez-Minaya}, \binits{J.}},
\oauthor{\bsnm{Paradinas}, \binits{I.}},
\oauthor{\bsnm{Pennino}, \binits{M.G.}}:
Bayesian feedback in the framework of ecological sciences
(2024)
{\href{https://arxiv.org/abs/2305.17922}{{arXiv:2305.17922}}}
{[stat.AP]}
\end{botherref}
\endbibitem

\bibitem[\protect\citeauthoryear{Rue et~al.}{2009}]{Rue_INLA_2009}
\begin{barticle}
\bauthor{\bsnm{Rue}, \binits{H.}},
\bauthor{\bsnm{Martino}, \binits{S.}},
\bauthor{\bsnm{Chopin}, \binits{N.}}:
\batitle{{Approximate Bayesian inference for latent Gaussian models by using
  integrated nested Laplace approximations}}.
\bjtitle{Journal of the Royal Statistical Society: Series B (Statistical
  Methodology)}
\bvolume{71}(\bissue{2}),
\bfpage{319}--\blpage{392}
(\byear{2009})
\doiurl{10.1111/j.1467-9868.2008.00700.x}
{\href{https://arxiv.org/abs/https://rss.onlinelibrary.wiley.com/doi/pdf/10.1111/j.1467-9868.2008.00700.x}{{https://rss.onlinelibrary.wiley.com/doi/pdf/10.1111/j.1467-9868.2008.00700.x}}}
\end{barticle}
\endbibitem

\bibitem[\protect\citeauthoryear{Banerjee
  et~al.}{2015}]{Banerjee_HierarchicalSpatialData_2015}
\begin{bbook}
\bauthor{\bsnm{Banerjee}, \binits{S.}},
\bauthor{\bsnm{Carlin}, \binits{B.P.}},
\bauthor{\bsnm{Gelfand}, \binits{A.E.}}:
\bbtitle{{H}ierarchical {M}odeling and {A}nalysis for {S}patial {D}ata},
\bedition{2ed.} edn.
\bsertitle{Chapman \& Hall/CRC Monographs on Statistics \& Applied
  Probability}.
\bpublisher{Chapman and Hall/CRC},
\blocation{New York}
(\byear{2015}).
\doiurl{10.1201/b17115}
\end{bbook}
\endbibitem

\bibitem[\protect\citeauthoryear{Paradinas
  et~al.}{2017}]{Paradinas_SpatioTemporal_2017}
\begin{barticle}
\bauthor{\bsnm{Paradinas}, \binits{I.}},
\bauthor{\bsnm{Conesa}, \binits{D.}},
\bauthor{\bsnm{López-Quílez}, \binits{A.}},
\bauthor{\bsnm{Bellido}, \binits{J.M.}}:
\batitle{{Spatio-Temporal model structures with shared components for
  semi-continuous species distribution modelling}}.
\bjtitle{Spatial Statistics}
\bvolume{22},
\bfpage{434}--\blpage{450}
(\byear{2017})
\doiurl{10.1016/j.spasta.2017.08.001}
\end{barticle}
\endbibitem

\bibitem[\protect\citeauthoryear{Diggle
  et~al.}{1998}]{Diggle_Geostatistics_1998}
\begin{barticle}
\bauthor{\bsnm{Diggle}, \binits{P.J.}},
\bauthor{\bsnm{Tawn}, \binits{J.A.}},
\bauthor{\bsnm{Moyeed}, \binits{R.A.}}:
\batitle{{M}odel-{B}ased {G}eostatistics}.
\bjtitle{Journal of the Royal Statistical Society. Series C (Applied
  Statistics)}
\bvolume{47}(\bissue{3}),
\bfpage{299}--\blpage{350}
(\byear{1998})
\doiurl{10.1111/1467-9876.00113}
\end{barticle}
\endbibitem

\bibitem[\protect\citeauthoryear{Diggle and
  Ribeiro}{2007}]{Diggle_ModelBasedGeostatistics_2007}
\begin{bbook}
\bauthor{\bsnm{Diggle}, \binits{P.J.}},
\bauthor{\bsnm{Ribeiro}, \binits{P.J.}}:
\bbtitle{Model-Based Geostatistics}.
\bpublisher{Springer},
\blocation{New York}
(\byear{2007}).
\doiurl{10.1007/978-0-387-48536-2}
\end{bbook}
\endbibitem

\bibitem[\protect\citeauthoryear{Nychka
  et~al.}{2015}]{Nychka_MultiresolutionGaussianProcess_2015}
\begin{barticle}
\bauthor{\bsnm{Nychka}, \binits{D.}},
\bauthor{\bsnm{Bandyopadhyay}, \binits{S.}},
\bauthor{\bsnm{Hammerling}, \binits{D.}},
\bauthor{\bsnm{Lindgren}, \binits{F.}},
\bauthor{\bsnm{Sain}, \binits{S.}}:
\batitle{{A Multiresolution Gaussian Process Model for the Analysis of Large
  Spatial Datasets}}.
\bjtitle{Journal of Computational and Graphical Statistics}
\bvolume{24}(\bissue{2}),
\bfpage{579}--\blpage{599}
(\byear{2015})
\doiurl{10.1080/10618600.2014.914946} .
Accessed 2024-05-06
\end{barticle}
\endbibitem

\bibitem[\protect\citeauthoryear{Blangiardo
  et~al.}{2013}]{Blangiardo_SpatioTemporalINLA_2013}
\begin{barticle}
\bauthor{\bsnm{Blangiardo}, \binits{M.}},
\bauthor{\bsnm{Cameletti}, \binits{M.}},
\bauthor{\bsnm{Baio}, \binits{G.}},
\bauthor{\bsnm{Rue}, \binits{H.}}:
\batitle{{Spatial and spatio-temporal models with R-INLA}}.
\bjtitle{Spatial and Spatio-temporal Epidemiology}
\bvolume{7},
\bfpage{39}--\blpage{55}
(\byear{2013})
\doiurl{10.1016/j.sste.2013.07.003}
\end{barticle}
\endbibitem

\bibitem[\protect\citeauthoryear{Cosandey-Godin
  et~al.}{2015}]{Cosandey_SpatioTemporalArtic_2015}
\begin{barticle}
\bauthor{\bsnm{Cosandey-Godin}, \binits{A.}},
\bauthor{\bsnm{Krainski}, \binits{E.T.}},
\bauthor{\bsnm{Worm}, \binits{B.}},
\bauthor{\bsnm{Flemming}, \binits{J.M.}}:
\batitle{{Applying Bayesian spatiotemporal models to fisheries bycatch in the
  Canadian Arctic}}.
\bjtitle{Canadian Journal of Fisheries and Aquatic Sciences}
\bvolume{72}(\bissue{2}),
\bfpage{186}--\blpage{197}
(\byear{2015})
\doiurl{10.1139/cjfas-2014-0159}
{\href{https://arxiv.org/abs/https://doi.org/10.1139/cjfas-2014-0159}{{https://doi.org/10.1139/cjfas-2014-0159}}}
\end{barticle}
\endbibitem

\bibitem[\protect\citeauthoryear{Paradinas
  et~al.}{2015}]{Paradinas_SpatioTemporal_2015}
\begin{barticle}
\bauthor{\bsnm{Paradinas}, \binits{I.I.}},
\bauthor{\bsnm{Conesa}, \binits{D.}},
\bauthor{\bsnm{Pennino}, \binits{M.G.}},
\bauthor{\bsnm{Mu{\~{n}}oz}, \binits{F.}},
\bauthor{\bsnm{Fern{\'a}ndez}, \binits{A.M.}},
\bauthor{\bsnm{L{\'o}pez-Qu{\'i}lez}, \binits{A.}},
\bauthor{\bsnm{Bellido}, \binits{J.M.}}:
\batitle{{Bayesian spatio-temporal approach to identifying fish nurseries by
  validating persistence areas}}.
\bjtitle{Marine Ecology Progress Series}
\bvolume{528},
\bfpage{245}--\blpage{255}
(\byear{2015})
\doiurl{10.3354/meps11281}
\end{barticle}
\endbibitem

\bibitem[\protect\citeauthoryear{Huang
  et~al.}{2017}]{Jingyi_EnvironmentalINLA_2017}
\begin{barticle}
\bauthor{\bsnm{Huang}, \binits{J.}},
\bauthor{\bsnm{Malone}, \binits{B.P.}},
\bauthor{\bsnm{Minasny}, \binits{B.}},
\bauthor{\bsnm{McBratney}, \binits{A.B.}},
\bauthor{\bsnm{Triantafilis}, \binits{J.}}:
\batitle{{Evaluating a Bayesian modelling approach (INLA-SPDE) for
  environmental mapping}}.
\bjtitle{Science of The Total Environment}
\bvolume{609},
\bfpage{621}--\blpage{632}
(\byear{2017})
\doiurl{10.1016/j.scitotenv.2017.07.201}
\end{barticle}
\endbibitem

\bibitem[\protect\citeauthoryear{Diggle
  et~al.}{2010}]{Diggle_Preferential_2010}
\begin{barticle}
\bauthor{\bsnm{Diggle}, \binits{P.J.}},
\bauthor{\bsnm{Menezes}, \binits{R.}},
\bauthor{\bsnm{Su}, \binits{T.-l.}}:
\batitle{Geostatistical inference under preferential sampling}.
\bjtitle{Journal of the Royal Statistical Society: Series C (Applied
  Statistics)}
\bvolume{59}(\bissue{2}),
\bfpage{191}--\blpage{232}
(\byear{2010})
\doiurl{10.1111/j.1467-9876.2009.00701.x}
{\href{https://arxiv.org/abs/https://rss.onlinelibrary.wiley.com/doi/pdf/10.1111/j.1467-9876.2009.00701.x}{{https://rss.onlinelibrary.wiley.com/doi/pdf/10.1111/j.1467-9876.2009.00701.x}}}
\end{barticle}
\endbibitem

\bibitem[\protect\citeauthoryear{Illian et~al.}{2012}]{Illian_ToolboxLGCP_2012}
\begin{barticle}
\bauthor{\bsnm{Illian}, \binits{J.B.}},
\bauthor{\bsnm{S{\o}rbye}, \binits{S.H.}},
\bauthor{\bsnm{Rue}, \binits{H.}}:
\batitle{{A toolbox for fitting complex spatial point process models using
  integrated nested Laplace approximation (INLA)}}.
\bjtitle{The Annals of Applied Statistics}
\bvolume{6}(\bissue{4}),
\bfpage{1499}--\blpage{1530}
(\byear{2012})
\doiurl{10.1214/11-AOAS530}
\end{barticle}
\endbibitem

\bibitem[\protect\citeauthoryear{Diggle
  et~al.}{2013}]{Diggle_LogGaussianCoxProcesses_2013}
\begin{barticle}
\bauthor{\bsnm{Diggle}, \binits{P.J.}},
\bauthor{\bsnm{Moraga}, \binits{P.}},
\bauthor{\bsnm{Rowlingson}, \binits{B.}},
\bauthor{\bsnm{Taylor}, \binits{B.M.}}:
\batitle{{Spatial and Spatio-Temporal Log-Gaussian Cox Processes: Extending the
  Geostatistical Paradigm}}.
\bjtitle{Statistical Science}
\bvolume{28}(\bissue{4}),
\bfpage{542}--\blpage{563}
(\byear{2013})
\doiurl{10.1214/13-STS441}
\end{barticle}
\endbibitem

\bibitem[\protect\citeauthoryear{Simpson
  et~al.}{2016}]{Simpson_GoingOffGrid_2016}
\begin{barticle}
\bauthor{\bsnm{Simpson}, \binits{D.}},
\bauthor{\bsnm{Illian}, \binits{J.B.}},
\bauthor{\bsnm{Lindgren}, \binits{F.}},
\bauthor{\bsnm{Sørbye}, \binits{S.H.}},
\bauthor{\bsnm{Rue}, \binits{H.}}:
\batitle{{Going off grid: computationally efficient inference for log-Gaussian
  Cox processes}}.
\bjtitle{Biometrika}
\bvolume{103}(\bissue{1}),
\bfpage{49}--\blpage{70}
(\byear{2016})
\doiurl{10.1093/biomet/asv064}
\end{barticle}
\endbibitem

\bibitem[\protect\citeauthoryear{S{\o}rbye
  et~al.}{2018}]{Sorbye_priorLGCP_2018}
\begin{barticle}
\bauthor{\bsnm{S{\o}rbye}, \binits{S.H.}},
\bauthor{\bsnm{Illian}, \binits{J.B.}},
\bauthor{\bsnm{Simpson}, \binits{D.P.}},
\bauthor{\bsnm{Burslem}, \binits{D.}},
\bauthor{\bsnm{Rue}, \binits{H.}}:
\batitle{{Careful Prior Specification Avoids Incautious Inference for
  Log-Gaussian Cox Point Processes}}.
\bjtitle{Journal of the Royal Statistical Society Series C: Applied Statistics}
\bvolume{68}(\bissue{3}),
\bfpage{543}--\blpage{564}
(\byear{2018})
\doiurl{10.1111/rssc.12321}
\end{barticle}
\endbibitem

\bibitem[\protect\citeauthoryear{Krainski
  et~al.}{2018}]{Krainski_AdvancedSPDE_2018}
\begin{bbook}
\bauthor{\bsnm{Krainski}, \binits{E.}},
\bauthor{\bsnm{Gómez~Rubio}, \binits{V.}},
\bauthor{\bsnm{Bakka}, \binits{H.}},
\bauthor{\bsnm{Lenzi}, \binits{A.}},
\bauthor{\bsnm{Castro-Camilo}, \binits{D.}},
\bauthor{\bsnm{Simpson}, \binits{D.}},
\bauthor{\bsnm{Lindgren}, \binits{F.}},
\bauthor{\bsnm{Rue}, \binits{H.}}:
\bbtitle{{A}dvanced {S}patial {M}odeling with {S}tochastic {P}artial
  {D}ifferential {E}quations Using {R} and {INLA}}.
\bpublisher{Chapman and Hall/CRC Press},
\blocation{New York}
(\byear{2018}).
\doiurl{10.1201/9780429031892}
\end{bbook}
\endbibitem

\bibitem[\protect\citeauthoryear{Wikle
  et~al.}{2019}]{Wikle_SpatioTemporalStatistics_2019}
\begin{bbook}
\bauthor{\bsnm{Wikle}, \binits{C.}},
\bauthor{\bsnm{Mangion}, \binits{A.Z.}},
\bauthor{\bsnm{Cressie}, \binits{N.}}:
\bbtitle{Spatio-Temporal Statistics With R}.
\bpublisher{Chapman and Hall/CRC},
\blocation{New York}
(\byear{2019}).
\doiurl{10.1201/9781351769723}
\end{bbook}
\endbibitem

\bibitem[\protect\citeauthoryear{Moraga et~al.}{2021}]{Moraga_MalariaINLA_2021}
\begin{barticle}
\bauthor{\bsnm{Moraga}, \binits{P.}},
\bauthor{\bsnm{Dean}, \binits{C.}},
\bauthor{\bsnm{Inoue}, \binits{J.}},
\bauthor{\bsnm{Morawiecki}, \binits{P.}},
\bauthor{\bsnm{Noureen}, \binits{S.R.}},
\bauthor{\bsnm{Wang}, \binits{F.}}:
\batitle{{Bayesian spatial modelling of geostatistical data using INLA and SPDE
  methods: A case study predicting malaria risk in Mozambique}}.
\bjtitle{Spatial and Spatio-temporal Epidemiology}
\bvolume{39},
\bfpage{100440}
(\byear{2021})
\doiurl{10.1016/j.sste.2021.100440}
\end{barticle}
\endbibitem

\bibitem[\protect\citeauthoryear{Yuan
  et~al.}{2017}]{Yuan_PointProcessDistanceSampling_2017}
\begin{barticle}
\bauthor{\bsnm{Yuan}, \binits{Y.}},
\bauthor{\bsnm{Bachl}, \binits{F.E.}},
\bauthor{\bsnm{Lindgren}, \binits{F.}},
\bauthor{\bsnm{Borchers}, \binits{D.L.}},
\bauthor{\bsnm{Illian}, \binits{J.B.}},
\bauthor{\bsnm{Buckland}, \binits{S.T.}},
\bauthor{\bsnm{Rue}, \binits{H.}},
\bauthor{\bsnm{Gerrodette}, \binits{T.}}:
\batitle{{Point process models for spatio-temporal distance sampling data from
  a large-scale survey of blue whales}}.
\bjtitle{The Annals of Applied Statistics}
\bvolume{11}(\bissue{4}),
\bfpage{2270}--\blpage{2297}
(\byear{2017})
\doiurl{10.1214/17-AOAS1078}
\end{barticle}
\endbibitem

\bibitem[\protect\citeauthoryear{Arce~Guillen
  et~al.}{2023}]{Arce_SpatialVariationStepSelection_2023}
\begin{barticle}
\bauthor{\bsnm{Arce~Guillen}, \binits{R.}},
\bauthor{\bsnm{Lindgren}, \binits{F.}},
\bauthor{\bsnm{Muff}, \binits{S.}},
\bauthor{\bsnm{Glass}, \binits{T.W.}},
\bauthor{\bsnm{Breed}, \binits{G.A.}},
\bauthor{\bsnm{Schlägel}, \binits{U.E.}}:
\batitle{Accounting for unobserved spatial variation in step selection analyses
  of animal movement via spatial random effects}.
\bjtitle{Methods in Ecology and Evolution}
\bvolume{14}(\bissue{10}),
\bfpage{2639}--\blpage{2653}
(\byear{2023})
\doiurl{10.1111/2041-210X.14208}
{\href{https://arxiv.org/abs/https://besjournals.onlinelibrary.wiley.com/doi/pdf/10.1111/2041-210X.14208}{{https://besjournals.onlinelibrary.wiley.com/doi/pdf/10.1111/2041-210X.14208}}}
\end{barticle}
\endbibitem

\bibitem[\protect\citeauthoryear{Michelot
  et~al.}{2024}]{Michelot_StepSelectionAnalysis_2024}
\begin{barticle}
\bauthor{\bsnm{Michelot}, \binits{T.}},
\bauthor{\bsnm{Klappstein}, \binits{N.J.}},
\bauthor{\bsnm{Potts}, \binits{J.R.}},
\bauthor{\bsnm{Fieberg}, \binits{J.}}:
\batitle{{Understanding step selection analysis through numerical
  integration}}.
\bjtitle{Methods in Ecology and Evolution}
\bvolume{15}(\bissue{1}),
\bfpage{24}--\blpage{35}
(\byear{2024})
\doiurl{10.1111/2041-210X.14248}
{\href{https://arxiv.org/abs/https://besjournals.onlinelibrary.wiley.com/doi/pdf/10.1111/2041-210X.14248}{{https://besjournals.onlinelibrary.wiley.com/doi/pdf/10.1111/2041-210X.14248}}}
\end{barticle}
\endbibitem

\bibitem[\protect\citeauthoryear{Gómez-Rubio
  et~al.}{2021}]{Virgilio_SpatialEconometricsINLA_2021}
\begin{botherref}
\oauthor{\bsnm{Gómez-Rubio}, \binits{V.}},
\oauthor{\bsnm{Bivand}, \binits{R.S.}},
\oauthor{\bsnm{Rue}, \binits{H.}}:
{Estimating Spatial Econometrics Models with Integrated Nested Laplace
  Approximation}.
Mathematics
\textbf{9}(2044)
(2021)
\doiurl{10.3390/math9172044}
\end{botherref}
\endbibitem

\bibitem[\protect\citeauthoryear{Knorr-Held}{2000}]{KnorrHeld_SpaceTime_2000}
\begin{barticle}
\bauthor{\bsnm{Knorr-Held}, \binits{L.}}:
\batitle{{Bayesian modelling of inseparable space-time variation in disease
  risk}}.
\bjtitle{Statistics in Medicine}
\bvolume{19}(\bissue{17-18}),
\bfpage{2555}--\blpage{2567}
(\byear{2000})
\doiurl{10.1002/1097-0258(20000915/30)19:17/18<2555::AID-SIM587>3.0.CO;2-\#}
\end{barticle}
\endbibitem

\bibitem[\protect\citeauthoryear{Clayton}{1996}]{Clayton_GLMM_1996}
\begin{bchapter}
\bauthor{\bsnm{Clayton}, \binits{D.}}:
\bctitle{{Generalized linear mixed models}}.
In: \beditor{\bsnm{Gilks}, \binits{W.R.}},
\beditor{\bsnm{Richardson}, \binits{S.}},
\beditor{\bsnm{Spiegelhalter}, \binits{D.J.}} (eds.)
\bbtitle{Markov Chain Monte Carlo in Practice},
pp. \bfpage{275}--\blpage{301}.
\bpublisher{Chapman and Hall},
\blocation{London}
(\byear{1996}).
\bcomment{Chap. 16}
\end{bchapter}
\endbibitem

\bibitem[\protect\citeauthoryear{Lindgren
  et~al.}{2024}]{Lindgren_diffusionbased_2024}
\begin{barticle}
\bauthor{\bsnm{Lindgren}, \binits{F.}},
\bauthor{\bsnm{Bakka}, \binits{H.}},
\bauthor{\bsnm{Bolin}, \binits{D.}},
\bauthor{\bsnm{Krainski}, \binits{E.}},
\bauthor{\bsnm{Rue}, \binits{H.}}:
\batitle{{A diffusion-based spatio-temporal extension of Gaussian Mat\'ern
  fields}}.
\bjtitle{SORT-Statistics and Operations Research Transactions}
\bvolume{48}(\bissue{1}),
\bfpage{3}--\blpage{66}
(\byear{2024})
\doiurl{10.57645/20.8080.02.13}
\end{barticle}
\endbibitem

\bibitem[\protect\citeauthoryear{Koshkina
  et~al.}{2017}]{Koshkina_ISDMimperfectdetection_2017}
\begin{barticle}
\bauthor{\bsnm{Koshkina}, \binits{V.}},
\bauthor{\bsnm{Wang}, \binits{Y.}},
\bauthor{\bsnm{Gordon}, \binits{A.}},
\bauthor{\bsnm{Dorazio}, \binits{R.M.}},
\bauthor{\bsnm{White}, \binits{M.}},
\bauthor{\bsnm{Stone}, \binits{L.}}:
\batitle{{Integrated species distribution models: combining presence-background
  data and site-occupancy data with imperfect detection}}.
\bjtitle{Methods in Ecology and Evolution}
\bvolume{8}(\bissue{4}),
\bfpage{420}--\blpage{430}
(\byear{2017})
\doiurl{10.1111/2041-210X.12738}
{\href{https://arxiv.org/abs/https://besjournals.onlinelibrary.wiley.com/doi/pdf/10.1111/2041-210X.12738}{{https://besjournals.onlinelibrary.wiley.com/doi/pdf/10.1111/2041-210X.12738}}}
\end{barticle}
\endbibitem

\bibitem[\protect\citeauthoryear{Jung}{2023}]{Jung_IntegratedSpeciesDistributionModels_2023}
\begin{barticle}
\bauthor{\bsnm{Jung}, \binits{M.}}:
\batitle{{An integrated species distribution modelling framework for
  heterogeneous biodiversity data}}.
\bjtitle{Ecological Informatics}
\bvolume{76},
\bfpage{102127}
(\byear{2023})
\doiurl{10.1016/j.ecoinf.2023.102127}
\end{barticle}
\endbibitem

\bibitem[\protect\citeauthoryear{Martins
  et~al.}{2013}]{Martins_BayesianComputingINLA_2013}
\begin{barticle}
\bauthor{\bsnm{Martins}, \binits{T.G.}},
\bauthor{\bsnm{Simpson}, \binits{D.}},
\bauthor{\bsnm{Lindgren}, \binits{F.}},
\bauthor{\bsnm{Rue}, \binits{H.}}:
\batitle{{Bayesian computing with INLA: new features}}.
\bjtitle{Computational Statistics \& Data Analysis}
\bvolume{67},
\bfpage{68}--\blpage{83}
(\byear{2013})
\doiurl{10.1016/j.csda.2013.04.014}
\end{barticle}
\endbibitem

\bibitem[\protect\citeauthoryear{Rue and Leonhard}{2005}]{Rue_GMRF_2005}
\begin{bbook}
\bauthor{\bsnm{Rue}, \binits{H.}},
\bauthor{\bsnm{Leonhard}, \binits{H.}}:
\bbtitle{Gaussian Markov Random Fields}.
\bpublisher{Chapman and Hall/CRC},
\blocation{New York}
(\byear{2005}).
\doiurl{10.1201/9780203492024}
\end{bbook}
\endbibitem

\bibitem[\protect\citeauthoryear{{Van Niekerk}
  et~al.}{2023}]{VanNiekert_NewINLA_2023}
\begin{barticle}
\bauthor{\bsnm{{Van Niekerk}}, \binits{J.}},
\bauthor{\bsnm{Krainski}, \binits{E.}},
\bauthor{\bsnm{Rustand}, \binits{D.}},
\bauthor{\bsnm{Rue}, \binits{H.}}:
\batitle{{A new avenue for Bayesian inference with INLA}}.
\bjtitle{Computational Statistics \& Data Analysis}
\bvolume{181},
\bfpage{107692}
(\byear{2023})
\doiurl{10.1016/j.csda.2023.107692}
\end{barticle}
\endbibitem

\bibitem[\protect\citeauthoryear{Spiegelhalter
  et~al.}{2002}]{Spiegelhalter_DIC_2002}
\begin{barticle}
\bauthor{\bsnm{Spiegelhalter}, \binits{D.J.}},
\bauthor{\bsnm{Best}, \binits{N.G.}},
\bauthor{\bsnm{Carlin}, \binits{B.P.}},
\bauthor{\bsnm{Van Der~Linde}, \binits{A.}}:
\batitle{{Bayesian measures of model complexity and fit}}.
\bjtitle{Journal of the Royal Statistical Society: Series B (Statistical
  Methodology)}
\bvolume{64}(\bissue{4}),
\bfpage{583}--\blpage{639}
(\byear{2002})
\doiurl{10.1111/1467-9868.00353}
{\href{https://arxiv.org/abs/https://rss.onlinelibrary.wiley.com/doi/pdf/10.1111/1467-9868.00353}{{https://rss.onlinelibrary.wiley.com/doi/pdf/10.1111/1467-9868.00353}}}
\end{barticle}
\endbibitem

\bibitem[\protect\citeauthoryear{Watanabe}{2013}]{Watanabe_WAIC_2013}
\begin{barticle}
\bauthor{\bsnm{Watanabe}, \binits{S.}}:
\batitle{{A widely applicable Bayesian information criterion}}.
\bjtitle{J. Mach. Learn. Res.}
\bvolume{14}(\bissue{1}),
\bfpage{867}--\blpage{897}
(\byear{2013})
\end{barticle}
\endbibitem

\bibitem[\protect\citeauthoryear{Pettit}{1990}]{Pettit_CPO_1990}
\begin{barticle}
\bauthor{\bsnm{Pettit}, \binits{L.I.}}:
\batitle{{The Conditional Predictive Ordinate for the Normal Distribution}}.
\bjtitle{Journal of the Royal Statistical Society: Series B (Methodological)}
\bvolume{52}(\bissue{1}),
\bfpage{175}--\blpage{184}
(\byear{1990})
\doiurl{10.1111/j.2517-6161.1990.tb01780.x}
{\href{https://arxiv.org/abs/https://rss.onlinelibrary.wiley.com/doi/pdf/10.1111/j.2517-6161.1990.tb01780.x}{{https://rss.onlinelibrary.wiley.com/doi/pdf/10.1111/j.2517-6161.1990.tb01780.x}}}
\end{barticle}
\endbibitem

\bibitem[\protect\citeauthoryear{Lindgren
  et~al.}{2011}]{Lindgren_ExplicitLinkSPDE_2011}
\begin{barticle}
\bauthor{\bsnm{Lindgren}, \binits{F.}},
\bauthor{\bsnm{Rue}, \binits{H.}},
\bauthor{\bsnm{Lindström}, \binits{J.}}:
\batitle{{An explicit link between Gaussian fields and Gaussian Markov random
  fields: the stochastic partial differential equation approach}}.
\bjtitle{Journal of the Royal Statistical Society: Series B (Statistical
  Methodology)}
\bvolume{73}(\bissue{4}),
\bfpage{423}--\blpage{498}
(\byear{2011})
\doiurl{10.1111/j.1467-9868.2011.00777.x}
{\href{https://arxiv.org/abs/https://rss.onlinelibrary.wiley.com/doi/pdf/10.1111/j.1467-9868.2011.00777.x}{{https://rss.onlinelibrary.wiley.com/doi/pdf/10.1111/j.1467-9868.2011.00777.x}}}
\end{barticle}
\endbibitem

\bibitem[\protect\citeauthoryear{Bakka
  et~al.}{2019}]{Bakka_SpatialBarriersINLA_2019}
\begin{barticle}
\bauthor{\bsnm{Bakka}, \binits{H.}},
\bauthor{\bsnm{Vanhatalo}, \binits{J.}},
\bauthor{\bsnm{Illian}, \binits{J.B.}},
\bauthor{\bsnm{Simpson}, \binits{D.}},
\bauthor{\bsnm{Rue}, \binits{H.}}:
\batitle{{Non-stationary Gaussian models with physical barriers}}.
\bjtitle{Spatial Statistics}
\bvolume{29},
\bfpage{268}--\blpage{288}
(\byear{2019})
\doiurl{10.1016/j.spasta.2019.01.002}
\end{barticle}
\endbibitem

\bibitem[\protect\citeauthoryear{Sid{\'e}n
  et~al.}{2021}]{Siden_3DSpatialMatern_2021}
\begin{barticle}
\bauthor{\bsnm{Sid{\'e}n}, \binits{P.}},
\bauthor{\bsnm{Lindgren}, \binits{F.}},
\bauthor{\bsnm{Bolin}, \binits{D.}},
\bauthor{\bsnm{Eklund}, \binits{A.}},
\bauthor{\bsnm{Villani}, \binits{M.}}:
\batitle{{Spatial 3D Matérn Priors for Fast Whole-Brain fMRI Analysis}}.
\bjtitle{Bayesian Analysis}
\bvolume{16}(\bissue{4}),
\bfpage{1251}--\blpage{1278}
(\byear{2021})
\doiurl{10.1214/21-BA1283}
\end{barticle}
\endbibitem

\bibitem[\protect\citeauthoryear{Pennino
  et~al.}{2019}]{Pennino_AccountingPreferential_2019}
\begin{barticle}
\bauthor{\bsnm{Pennino}, \binits{M.G.}},
\bauthor{\bsnm{Paradinas}, \binits{I.}},
\bauthor{\bsnm{Illian}, \binits{J.B.}},
\bauthor{\bsnm{Muñoz}, \binits{F.}},
\bauthor{\bsnm{Bellido}, \binits{J.M.}},
\bauthor{\bsnm{López-Quílez}, \binits{A.}},
\bauthor{\bsnm{Conesa}, \binits{D.}}:
\batitle{{Accounting for preferential sampling in species distribution
  models}}.
\bjtitle{Ecology and Evolution}
\bvolume{9}(\bissue{1}),
\bfpage{653}--\blpage{663}
(\byear{2019})
\doiurl{10.1002/ece3.4789}
{\href{https://arxiv.org/abs/https://onlinelibrary.wiley.com/doi/pdf/10.1002/ece3.4789}{{https://onlinelibrary.wiley.com/doi/pdf/10.1002/ece3.4789}}}
\end{barticle}
\endbibitem

\bibitem[\protect\citeauthoryear{Lezama-Ochoa
  et~al.}{2020}]{Nerea_SpinetailSPDE_2020}
\begin{barticle}
\bauthor{\bsnm{Lezama-Ochoa}, \binits{N.}},
\bauthor{\bsnm{Pennino}, \binits{M.G.}},
\bauthor{\bsnm{Hall}, \binits{M.A.}},
\bauthor{\bsnm{Lopez}, \binits{J.}},
\bauthor{\bsnm{Murua}, \binits{H.}}:
\batitle{{Using a Bayesian modelling approach (INLA-SPDE) to predict the
  occurrence of the Spinetail Devil Ray (Mobular mobular)}}.
\bjtitle{Scientific Reports}
\bvolume{10}(\bissue{1}),
\bfpage{18822}
(\byear{2020})
\doiurl{10.1038/s41598-020-73879-3}
\end{barticle}
\endbibitem

\bibitem[\protect\citeauthoryear{Bivand
  et~al.}{2014}]{Bivand_SpatialEconometricsINLA_2014}
\begin{barticle}
\bauthor{\bsnm{Bivand}, \binits{R.S.}},
\bauthor{\bsnm{Gómez-Rubio}, \binits{V.}},
\bauthor{\bsnm{Rue}, \binits{H.}}:
\batitle{{Approximate Bayesian inference for spatial econometrics models}}.
\bjtitle{Spatial Statistics}
\bvolume{9},
\bfpage{146}--\blpage{165}
(\byear{2014})
\doiurl{10.1016/j.spasta.2014.01.002}
\end{barticle}
\endbibitem

\bibitem[\protect\citeauthoryear{Huang and
  Gelman}{2005}]{Huang_SamplingFB_2005}
\begin{botherref}
\oauthor{\bsnm{Huang}, \binits{Z.}},
\oauthor{\bsnm{Gelman}, \binits{A.}}:
{Sampling for Bayesian Computation with Large Datasets}.
Technical report
(2005).
\doiurl{10.2139/ssrn.1010107}
\end{botherref}
\endbibitem

\bibitem[\protect\citeauthoryear{Hayya
  et~al.}{1975}]{Hayya_NoteRatioTwoGaussians_1975}
\begin{barticle}
\bauthor{\bsnm{Hayya}, \binits{J.}},
\bauthor{\bsnm{Armstrong}, \binits{D.}},
\bauthor{\bsnm{Gressis}, \binits{N.}}:
\batitle{{A Note on the Ratio of Two Normally Distributed Variables}}.
\bjtitle{Management Science}
\bvolume{21}(\bissue{11}),
\bfpage{1338}--\blpage{1341}
(\byear{1975})
\doiurl{10.1287/mnsc.21.11.1338}
{\href{https://arxiv.org/abs/https://doi.org/10.1287/mnsc.21.11.1338}{{https://doi.org/10.1287/mnsc.21.11.1338}}}
\end{barticle}
\endbibitem

\bibitem[\protect\citeauthoryear{Orozco-Acosta et~al.}{2023}]{Aritz_BigDM_2023}
\begin{barticle}
\bauthor{\bsnm{Orozco-Acosta}, \binits{E.}},
\bauthor{\bsnm{Adin}, \binits{A.}},
\bauthor{\bsnm{Ugarte}, \binits{M.D.}}:
\batitle{{Big problems in spatio-temporal disease mapping: Methods and
  software}}.
\bjtitle{Computer Methods and Programs in Biomedicine}
\bvolume{231},
\bfpage{107403}
(\byear{2023})
\doiurl{10.1016/j.cmpb.2023.107403}
\end{barticle}
\endbibitem

\bibitem[\protect\citeauthoryear{Vicente
  et~al.}{2023}]{Aritz_HighDimensionalMultivariate_2023}
\begin{barticle}
\bauthor{\bsnm{Vicente}, \binits{G.}},
\bauthor{\bsnm{Adin}, \binits{A.}},
\bauthor{\bsnm{Goicoa}, \binits{T.}},
\bauthor{\bsnm{Ugarte}, \binits{M.D.}}:
\batitle{{High-dimensional order-free multivariate spatial disease mapping}}.
\bjtitle{Statistics and Computing}
\bvolume{33}(\bissue{5}),
\bfpage{104}
(\byear{2023})
\doiurl{10.1007/s11222-023-10263-x}
\end{barticle}
\endbibitem

\end{thebibliography}

\end{document}